\documentclass[twocolumn]{aastex631}

\begin{document}
\title{Compact Binary Merger Rate in Dark-Matter Spikes}

\author[0000-0002-6349-8489]{Saeed Fakhry} 
\email{s\_fakhry@sbu.ac.ir}
\affiliation{Department of Physics, Shahid Beheshti University, Evin, Tehran 19839, Iran}
\affiliation{PDAT Laboratory, Department of Physics, K.N. Toosi University of Technology, P.O. Box 15875-4416, Tehran, Iran}
\author{Zahra Salehnia}
\email{zahra.salehnia@email.kntu.ac.ir}
\affiliation{Department of Physics, K.N. Toosi University of Technology, P.O. Box 15875-4416, Tehran, Iran}
\affiliation{PDAT Laboratory, Department of Physics, K.N. Toosi University of Technology, P.O. Box 15875-4416, Tehran, Iran}
\author{Azin Shirmohammadi}
\email{azinshr@email.kntu.ac.ir}
\affiliation{Department of Physics, K.N. Toosi University of Technology, P.O. Box 15875-4416, Tehran, Iran}
\affiliation{PDAT Laboratory, Department of Physics, K.N. Toosi University of Technology, P.O. Box 15875-4416, Tehran, Iran}
\author{Mina Ghodsi Yengejeh}
\email{m.ghodsi.y@gmail.com}
\affiliation{Department of Physics, Shahid Beheshti University, Evin, Tehran 19839, Iran}
\affiliation{PDAT Laboratory, Department of Physics, K.N. Toosi University of Technology, P.O. Box 15875-4416, Tehran, Iran}
\author{Javad T. Firouzjaee}
\email{firouzjaee@kntu.ac.ir}
\affiliation{Department of Physics, K.N. Toosi University of Technology, P.O. Box 15875-4416, Tehran, Iran}
\affiliation{School of Physics, Institute for Research in Fundamental Sciences (IPM), P.O. Box 19395-5531, Tehran, Iran}
\affiliation{PDAT Laboratory, Department of Physics, K.N. Toosi University of Technology, P.O. Box 15875-4416, Tehran, Iran}

\begin{abstract}
Nowadays, the existence of supermassive black holes (SMBHs) in the center of galactic halos is almost confirmed. An extremely dense region referred to as dark-matter spike is expected to form around central SMBHs as they grow and evolve adiabatically. In this work, we calculate the merger rate of compact binaries in dark-matter spikes while considering halo models with spherical and ellipsoidal collapses. Our findings exhibit that ellipsoidal-collapse dark matter halo models can potentially yield the enhancement of the merger rate of compact binaries. Finally, our results confirm that the merger rate of primordial black hole binaries is consistent with the results estimated by the LIGO-Virgo detectors, while such results can not be realized for primordial black hole-neutron star binaries.
\end{abstract}


\section{Introduction} \label{sec:intro}
Over the last few decades, gravitational waves (GWs) have been studied as an interesting cosmological observable. Regarding this, a large part of astrophysical and cosmological phenomena have been evaluated through the study of GWs and their direct detections. Meanwhile, compact binary mergers are always considered potential sources for the propagation of GWs \citep{2022LRR....25....1M}. During recent years, dozens of compact binary mergers have been recorded by the LIGO-Virgo detectors (e.g., \cite{2016PhRvL.116f1102A, 2016PhRvL.116x1103A, 2016PhRvL.116v1101A, 2020ApJ...896L..44A, 2020PhRvL.125j1102A}). In this regard, three main classes of compact binaries as binary black holes (BBHs), black hole-neutron star (BH-NS) binaries, and binary neutron stars (BNSs) can potentially be captured by the LIGO-Virgo detectors. Meanwhile, most of the recorded GW events belong to BBH merger events in the mass range $10\,M_{\odot}\mbox{-}\,100\,M_{\odot}$ \citep{2019PhRvX...9c1040A, 2021PhRvX..11b1053A, 2021arXiv211103606T}. There have been several investigations into the origin of such BHs (e.g., \cite{2019JCAP...02..018R, 2021MNRAS.507.5224B, 2021PhRvL.126b1103N, 2021PhRvL.127o1101N, 2022PhR...955....1M}). However, our information on this matter is still insufficient. There is a possibility that they have formed during stellar collapse (likely via different channels) or that they have formed at the beginning of the Universe as a result of gravitational collapse of density fluctuations.

Surprisingly, most of the BBH merger events reported by the LIGO-Virgo collaboration are consistent with the primordial black hole (PBH) scenario. In the cosmological perturbation theory, PBHs are predicted to form due to nonlinear peaks in initial density fluctuations during horizon reentry (see, e.g., \cite{1967SvA....10..602Z, 1971MNRAS.152...75H, 1974MNRAS.168..399C}). Several studies show that the existence of a critical state in primordial density fluctuations seems necessary for the formation of PBHs, which is achieved through exceeding a threshold value (e.g., \cite{1999PhRvD..60h4002S, 2007CQGra..24.1405P, 2013CQGra..30n5009M, 2014JCAP...07..045Y, 2015arXiv150402071B, 2017JCAP...06..041A}). According to this argument, exceeding a certain threshold value is equivalent to the direct collapse of primordial density fluctuations and consequently the formation of PBHs. In addition, PBHs are characterized by their broad range of masses, which makes them distinct from astrophysical black holes \citep{2018CQGra..35f3001S}. Meanwhile, the standard model of cosmology tries to explain the nature of two basic components of the dark sector of the Universe: dark matter and dark energy. In this regard, dark energy is considered a type of energy that governs the accelerating expansion of the late-time Universe, which corresponds to the cosmological constant in general relativity (see, e.g., \cite{2011PhRvL.107b1302S, 2014JCAP...08..034Y, 2018RPPh...81a6901H, 2020PDU....3000666D, 2021arXiv210700562G, 2023PDU....3901144G}). Dark matter is also attributed to an invisible matter that is thought to make up approximately $85\%$ of the matter in the Universe. Nowadays, it is believed that PBHs can be regarded as one of the potential candidates for dark matter (e.g., \cite{2016PhRvL.116t1301B, 2016PhRvL.117f1101S, 2016ApJ...823L..25K, 2016JCAP...11..036B, 2017PDU....15..142C, 2018JCAP...01..004B, 2021FrASS...8...87V, 2021arXiv211002821C}). However, other candidates for dark matter are being discussed seriously (e.g., \cite{2000PhRvL..84.3760S, 2000ApJ...542..281A, 2012LRR....15....6L, 2018RPPh...81f6201R, 2020PhRvL.124j1303B, 2020PhRvD.102h4063D, 2021CQGra..38s4001D}). Efficient observational methods have been used to constrain the abundance of PBHs in different mass ranges, which in turn are compelling contexts for studying the early Universe at small scales (e.g., \cite{2017PhRvD..96b3514C, 2018JCAP...04..007L, 2021RPPh...84k6902C}). However, it is possible to obtain strong constraints on the contribution of PBHs to dark matter by calculating their merger rate and validating the results via GW data \citep{2016PhRvL.116t1301B, 2017PDU....15..142C}.

Furthermore, the random distribution of BHs in the Universe allows them to form binaries with NSs. Interestingly, BH-NS mergers can provide important information about multi-messenger astronomy \citep{2021FrASS...8...39R}. In addition to emitting of GWs, they can also transmit electromagnetic signals during the merger process \citep{2020EPJA...56....8B}.  In such events, residual matter from the NS is usually accreted by the BH and results in a luminous event \citep{2021ApJ...923L...2F}. Moreover, recording and processing data from BH-NS binary mergers by GW detectors can include individual information about the NS nuclear equation of state and accretion processes of BHs, as well as constraining their spin and abundance (see, e.g., \cite{2019PhRvD.100f3021H, 2019PhRvL.123d1102Z, 2021ApJ...918L..38F, 2021PhRvD.104l3024T}). In recent years, two merger events have been captured by the LIGO-Virgo detectors, which are attributed to BH-NS binaries and their component masses are ($8.9^{+1.2}_{-1.5}M_{\odot}, 1.9^{+0.3}_{-0.2} M_{\odot}$) and ($5.7^{+1.8}_{-2.1}M_{\odot}, 1.5^{+0.7}_{-0.3}M_{\odot}$), respectively \cite{2021ApJ...915L...5A}. There are many uncertainties surrounding the formation of BH-NS binaries and their merging. Nevertheless, describing this class of merger events using the evolution of the field binaries can be a viable approach. It is expected that a significant number of BH-NS merger events can be detected by GW detectors in the upcoming years. Due to these circumstances, it seems imperative to fully understand how compact objects involved in these events are formed. In \cite{2022ApJ...931....2S}, taking into account spherical-collapse dark matter halo models and the framework of the PBH scenario, the merger rate of BH-NS binaries has been calculated. Also, by comparing the results with the GW observations, it has been claimed that the BH components participating in such events can not have primordial origins. However, by considering more realistic halo models (e.g., those with ellipsoidal-collapse), a more accurate picture of galactic halos can be obtained (e.g., \cite{2021PhRvD.103l3014F, 2022PhRvD.105d3525F, 2022arXiv221013558F, 2022arXiv221208646F}). With this argument, in \cite{2022ApJ...941...36F}, we have indicated that by considering ellipsoidal-collapse dark matter halo models and validating the results with relevant events estimated by the LIGO-Virgo detectors, the BHs contributing to the BH-NS events are most likely PBHs.

On the other hand, there is convincing evidence of the presence of supermassive black holes (SMBHs) in the center of galactic halos (see, e.g., \cite{2021NatRP...3..732V, 2021AstL...47..515P, 2022PhRvD.106d3018S, 2022MNRAS.511.5436D}). This can be deduced from the Keplerian behavior of the velocity dispersion of the stars in the inner regions of galactic halos \citep{1998ApJ...509..678G}. It is believed that central SMBHs are capable of amplifying the density of surrounding dark matter particles to a certain extent. In this regard, in \cite{1999PhRvL..83.1719G}, it has been proposed that a dense region known as the dark-matter spike will form around the central SMBH if it evolves adiabatically and initially has a power-law cusp. Given that the dark matter density is very high in the spike regions, it is expected that the number density of PBHs is also significant in such regions. It is noteworthy that the structure of dark-matter spikes is determined by the growth of the central SMBH and dark matter halo model. In \cite{2019PhRvD..99d3533N}, the merger rate of PBH-PBH binaries has been calculated in dark-matter spikes while accounting for spherical-collapse dark matter halo models. However, as mentioned earlier, more realistic halo models (e.g., those with ellipsoidal collapse) can potentially affect the final results.

In this work, we propose to employ the ellipsoidal-collapse dark matter halo models to calculate the merger rate of compact binaries within dark-matter spikes. In this regard, the outline of this work is as follows. In Sec.\,\ref{sec:ii}, we introduce a convenient model for dark-matter spikes and discuss some crucial quantities such as spike density profile, SMBH mass function, and concentration parameter. Next, in Sec.\,\ref{sec:iii}, we calculate the merger rate of compact binaries in the framework of ellipsoidal-collapse dark matter halo models and compare it with that obtained from spherical-collapse dark matter halo models. We also compare the relevant results of the present analysis with the corresponding data estimated by the LIGO-Virgo detectors and constrain the value of the power-law index. Finally, we scrutinize the results and summarize the findings in Sec.\,\ref{sec:iv}.
\section{Model of dark-matter spike} \label{sec:ii}
\subsection{The density profile}\label{sec.iia}
According to the standard model of cosmology, dark matter halos are known as nonlinear structures that have been formed hierarchically and distributed in the Universe as a result of the collapse of linear cosmological fluctuations. Based on indirect observations from the rotation curve of galaxies, it can be argued that dark matter particles should not be uniformly distributed in galactic halos. In this regard, special attention should be paid to the SMBHs structured at the galactic center. Although ordinary black holes might be born from a stellar collapse, SMBHs are difficult to accommodate in standard astrophysical scenarios at high redshifts. It is suggested that SMBH mass can be related to the mass of the dark matter halo, implying that SMBHs could coevolve with their host halos \citep{2021NatRP...3..732V, 2022arXiv220601443B}. Accordingly, self-interacting dark matter halo models predict early seeds for supermassive black holes through the gravothermal catastrophe \citep{2019JCAP...07..036C, 2021ApJ...914L..26F, 2022JCAP...08..032F}. Consequently, the SMBH is expected to be surrounded by a highly dense spike of dark matter at the center of the galactic halo.

Assume $M_{\rm SMBH}$ is the mass of SMBH at the galactic center, which takes a density profile of the form $\rho(r) \simeq \rho_0(r_0/r)^\gamma$, where $\rho_0$ and $r_0$ are characteristic parameters of the halo, and $\gamma$ represents the power-law index. In light of this reasoning, one can expect to form a dark-matter spike whose radius is specified by the following relation \citep{1999PhRvL..83.1719G}
\begin{equation}
r_{\rm sp} = a_{\gamma}r_{0}\left(\frac{M_{\rm SMBH}}{\rho_{0} r_{0}^{3}}\right)^{1/(3-\gamma)},
\end{equation}
where $a_{\gamma}$ is determined through numerical suggestions for each power-law index $\gamma$.

Accordingly, the density profile of the dark-matter spike for $r$ in the range of $4r_{\rm s}<r< r_{\rm sp}$ can be obtained as follows \citep{1999PhRvL..83.1719G, 2019PhRvD..99d3533N}
\begin{equation}
\rho_{\rm sp}(r) = \rho_{0}\left(\frac{r_{0}}{r_{\rm sp}}\right)^{\gamma}\left(1-\frac{4r_{\rm s}}{r}\right)^{3}\left(\frac{r_{\rm sp}}{r}\right)^{\gamma_{\rm sp}},
\end{equation}
where $\gamma_{\rm sp}=(9-2\gamma)/(4-\gamma)$. Furthermore, $r_{\rm s}$ indicates the Schwarzchild radius of SMBH which is in the form of
\begin{equation}
r_{\rm s} = \frac{2GM_{\rm SMBH}}{c^2} \simeq 2.95\,{\rm km}\, \left(\frac{M_{\rm SMBH}}{M_{\odot}}\right),
\end{equation}
in which $G$ is the gravitational constant and $c$ is the velocity of light in a vacuum. The results of numerical simulations and analytical approaches exhibit that the density profile in small radii has a power-law form \citep{2006AJ....132.2685M,2009MNRAS.398L..21S,2010MNRAS.402...21N}. However, there are different predictions about the value of the power-law index $\gamma$. In this work, we consider a range for the power-law index as $0<\gamma\leq 2$. On the other hand, to describe dark matter distribution in galactic halos, a convenient density profile was presented by Navarro, Frenk, and White (NFW) \citep{1996ApJ...462..563N}, which is specified by the following formula
\begin{equation}
\rho_{\rm NFW}(r)=\frac{\rho_0}{\left(r/r_{0}\right) \left(1+r/r_{0}\right)^2}.
\end{equation}

\begin{figure}[t] 
\includegraphics[width=0.45\textwidth]{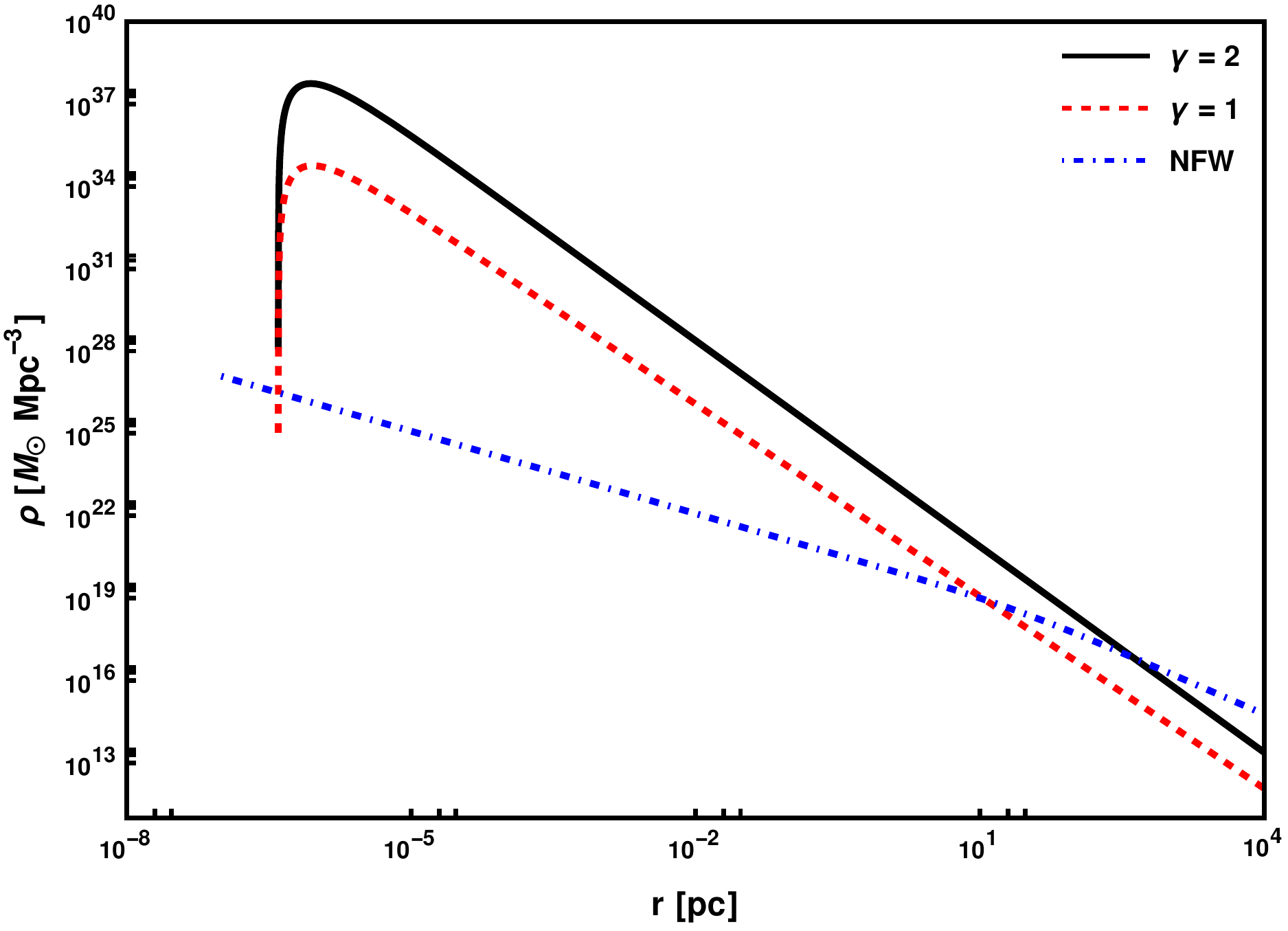}
\caption{Comparison of the NFW density profile with the profile of dark-matter spike with $\gamma=1$ and $2$ structured around the SMBH with a mass $M_{\rm SMBH}=10^{6}M_{\odot}$.}
\label{fig:1}
\end{figure}

In Fig.\,\ref{fig:1}, we have depicted the difference in the behavior of the density profile corresponding to the dark-matter spike with the NFW density profile while considering the mass of SMBH as $M_{\rm SMBH}=10^{6}M_{\odot}$. As can be seen from the figure, the density profile in the regions specific to the dark-matter spike is extremely higher than the NFW density profile. Therefore, it seems interesting to calculate the merger rate of compact binaries in dark-matter spikes. In addition, it should be noted that $r=r_{\rm sp}$ defines the radius within which the merger rate of compact objects must be calculated as dark-matter spike profiles cross the NFW density profile.
\subsection{The $M_{\rm SMBH}\mbox{-}\sigma$ relation}
Nowadays, there are several lines of evidence indicating that the growth of SMBHs and the evolution of their host halos are intricately linked (e.g., \cite{2002MNRAS.331..795M, 2002ApJ...574..740T, 2004ApJ...604L..89H, 2007MNRAS.378..198G, 2007ApJ...669...67H}). Regarding this, it is proposed that the mass of SMBH can be strongly correlated with the velocity dispersion of dark matter particles in a galactic halo, $\sigma$, which is called the $M_{\rm SMBH}\mbox{-}\sigma$ relation. In other words, the halo characteristic parameters, i.e., $\rho_0$ and $r_0$, can be related to $M_{\rm SMBH}$ by employing such a relation. A convenient form of the $M_{\rm SMBH}\mbox{-}\sigma$ relation has been obtained as \citep{2000ApJ...539L...9F, 2000ApJ...539L..13G}
\begin{equation} \label{masssmbh}
\log\left(\frac{M_{\rm SMBH}}{M_{\odot}}\right) =a+b \log\left(\frac{\sigma}{200\,{\rm km s^{-1}}}\right),
\end{equation}
where $a$ and $b$ are determined via empirical situation. In \cite{2009ApJ...698..198G}, values $a=8.12 \pm 0.08$ and $b=4.24 \pm 0.41$ have been suggested, which fits reasonably well with the various types of galactic halos \citep{2013ARA&A..51..511K}. The NFW profile is assumed to manage the density profile outside of spike regions, i.e. $r\gg r_{\rm sp}$, up to the virial radius $r_{\rm vir}>r_{0}$. In fact, $r_{\rm vir}$ is attributed to a radius that includes a volume in which the average halo density reaches $200$ to $500$ times the critical density of the Universe. Under such assumptions, the total mass enclosed by a volume of radius $r$ is determined as follows
\begin{equation}
M(r)=4 \pi \rho_{0} r_{0} \int_{0}^{r} \frac{r dr}{\left(1+r/r_{0}\right)^2}=4 \pi \rho_{0} r_{0}^{3} g(r/r_{0}),
\end{equation}
where $g(x)=\log(1+x)-x/(1+x)$. It needs to be mentioned that the contribution of dark-matter spike and central SMBH are negligible in comparison to the total mass of dark matter halo. Moreover, the concentration parameter determines the central density of dark matter halos, which is defined as $C\equiv r_{\rm vir}/r_{0}$. Hence, the virial mass takes the following form
\begin{equation}
M_{\rm vir}=4 \pi \rho_0 r_0^3 g(C).
\end{equation}
Also, the circular velocity of dark matter particles reaches the maximum value at a distance $r_{\rm m}=C_{\rm m}r_{0}=2.16\,r_{0}$, and corresponds to the one-dimensional velocity dispersion of dark matter particles, namely
\begin{equation} \label{sigmaaa}
\sigma^2 = \frac{GM(C_{\rm m}r_{0})}{C_{\rm m}r_{0}}=4\pi G\rho_0 r_0^2 \frac{g(C_{\rm m})}{C_{\rm m}}.
\end{equation}
As a result, a relation between $\rho_0$, $r_0$, and $M_{\rm SMBH}$ can be established via Eqs. (\ref{masssmbh}) and (\ref{sigmaaa}). Based on the results of $N$-body simulations, the concentration parameter is a decreasing function of halo mass and is a function of redshift at constant mass (e.g. \cite{2012MNRAS.423.3018P, 2014MNRAS.441.3359D, 2016MNRAS.460.1214L, 2016MNRAS.456.3068O}), which is consistent with the dynamics expected from the evolution of dark matter haloes. In this work, to calculate the merger rate of compact binaries in the present-time Universe, we utilize the concentration parameter presented in \cite{2016MNRAS.460.1214L} for spherical-collapse dark matter halo models, and we employ the corresponding one obtained in \cite{2016MNRAS.456.3068O} for ellipsoidal-collapse dark matter halo models.
\subsection{The mass function of SMBHs}
One of the most fundamental challenges of extragalactic astronomy is to understand how SMBHs grow and evolve. Accordingly, the SMBH mass function provides comprehensive information on the mass of SMBHs and their evolution at the center of galactic halos. Therefore, the SMBH mass function can be considered a powerful and available tool to investigate the growth of SMBHs and constrain related theoretical models. On the other hand, the SMBH mass function might play a significant role in the structuring of upcoming surveys because it provides an estimate of the mass classification of SMBHs \cite{2012AdAst2012E...7K}. It should be noted that obtaining an accurate mass function for SMBHs is a relatively difficult task. For this reason, the current estimates of the SMBH mass function include many theoretical uncertainties, which in turn may affect the accuracy of calculating the merger rate of compact binaries in dark-matter spikes. A reasonable approach for managing these uncertainties is to compare the results from several different empirical SMBH mass functions.

In \cite{2007MNRAS.379..841B}, by employing the {\it Galactica} code, a sample of $8839$ SDSS galaxies was employed to extrapolate the luminosity functions of spheroid and disc galaxies, and a mass function of SMBHs was obtained as follows
\begin{equation}\label{massfunc1}
\phi(M_{\rm SMBH}) = 10^{9}\left(\frac{\phi_{0} M_{\rm SMBH}^\alpha}{M_*^{\alpha+1}}\right)\exp\left[-\left(\frac{M_{\rm SMBH}}{M_*}\right)^{\beta}\right],
\end{equation}
in which $\alpha=-0.65$, $\beta = 0.6$, $\phi_{0}=2.9\times 10^{-3}\,h^3{\rm Mpc^{-3}}$, and $M_* = 4.07 \times 10^{7}\,h^{-2}M_{\odot}$.

In addition, in \cite{2009MNRAS.400.1451V}, a convenient mass function for SMBHs has been obtained via the {\it Millenium Galaxy Catalogue} \citep{2003MNRAS.344..307L} for $1743$ galaxies. This mass function is based on the experimental relation between the mass SMBH and the luminosity of the host spheroid, which has the following form
\begin{equation}\label{massfunc2}
\phi(M_{\rm SMBH}) = \phi_* \left(\frac{M_{\rm SMBH}}{M_*}\right)^{\alpha +1}\exp\left[ 1-\left(\frac{M_{\rm SMBH}}{M_*}\right)\right],
\end{equation}
where $\log\phi_*=-3.15\,h^{3}{\rm Mpc^{-3} dex^{-1}}$, $\log (M_*/M)=8.71$, and $\alpha=-1.20$. This mass function is valid for the masse range $10^6M_{\odot}<M_{\rm SMBH}< 10^{10}M_{\odot}$. 

Also, in \cite{2004MNRAS.354.1020S}, another suitable mass function was derived for SMBHs according to the observational relation between the SMBH mass and the halo velocity dispersion and using kinematic and photometric data, which takes the following formula
\begin{equation}\label{massfunc3}
\phi(M_{\rm SMBH}) = \phi_* \left(\frac{M_{\rm SMBH}}{M_*}\right)^{\alpha +1}\exp\left[ 1-\left(\frac{M_{\rm SMBH}}{M_*}\right)^\beta\right],
\end{equation}
where $\phi_*=7.7 \times 10^{-3}\,{\rm Mpc^{-3}}$, $M_*=6.4\times 10^{7}\,M_{\odot}$, $\beta=0.49$, and $\alpha=-1.11$. This mass function is valid for the mass range $10^{6}M_{\odot}\leq M_{\rm SMBH}\leq 5\times10^{9}M_{\odot}$.
\section{Compact binary merger rate} \label{sec:iii}
Assume in dark-matter spike, a compact object with mass $m_1$ suddenly encounters another one with mass $m_2$ on a hyperbolic orbit, and their relative velocity at large separation is $v_{\rm rel} = |v_1 - v_2|$. Hence, based on two-body scattering, highly significant gravitational radiation emits at the periastron $r_{\rm a}$. Keplerian mechanics states that such a system is gravitationally bound when the emitted gravitational energy dominates the kinetic energy of the system. Under these conditions, a maximum value for periastron can be obtained as follows
\begin{equation}\label{rmax}
r_{\rm mp}=\left[\frac{85\pi}{6\sqrt{2}}\frac{G^{7/2}m_{1}m_{2}(m_{1}+m_{2})^{3/2}}{c^{5}v_{\rm rel}^{2}}\right]^{2/7}.
\end{equation}
Furthermore, in the Newtonian limit, the impact parameter is determined by the periastron as follows:
\begin{equation}
b^{2}(r_{\rm p})=\frac{2G(m_1+m_2)r_{\rm p}}{v_{\rm rel}^{2}}+r_{\rm p}^{2}.
\end{equation}
For the regions of dark-matter spikes that are gravitationally active, a strong limit of gravitational focusing, i.e., $r_{\rm p}\ll b$, can be considered in such a way that the relevant distortions of surrounding compact objects on the formed binaries can be ignored. Thus, the cross-section for the binary formation can be obtained via the following equation
\begin{equation}\label{croossec}
\xi(m_1,m_2, v_{\rm rel})=\pi b^{2}(r_{\rm mp})\simeq \frac{2\pi G(m_1+m_2)r_{\rm mp}}{v_{\rm rel}^{2}}.
\end{equation}
Hence, by Substituting Eq.\,(\ref{rmax}) into Eq.\,(\ref{croossec}), the cross section for the binary formation can be derived as
\begin{equation}
\xi \simeq 2\pi \left(\frac{85\pi}{6\sqrt{2}}\right)^{2/7}\frac{G^{2}(m_1+m_2)^{10/7}(m_1m_2)^{2/7}}{c^{10/7}v_{\rm rel}^{18/7}}.
\end{equation}

\begin{figure*}[ht!]
\gridline{\fig{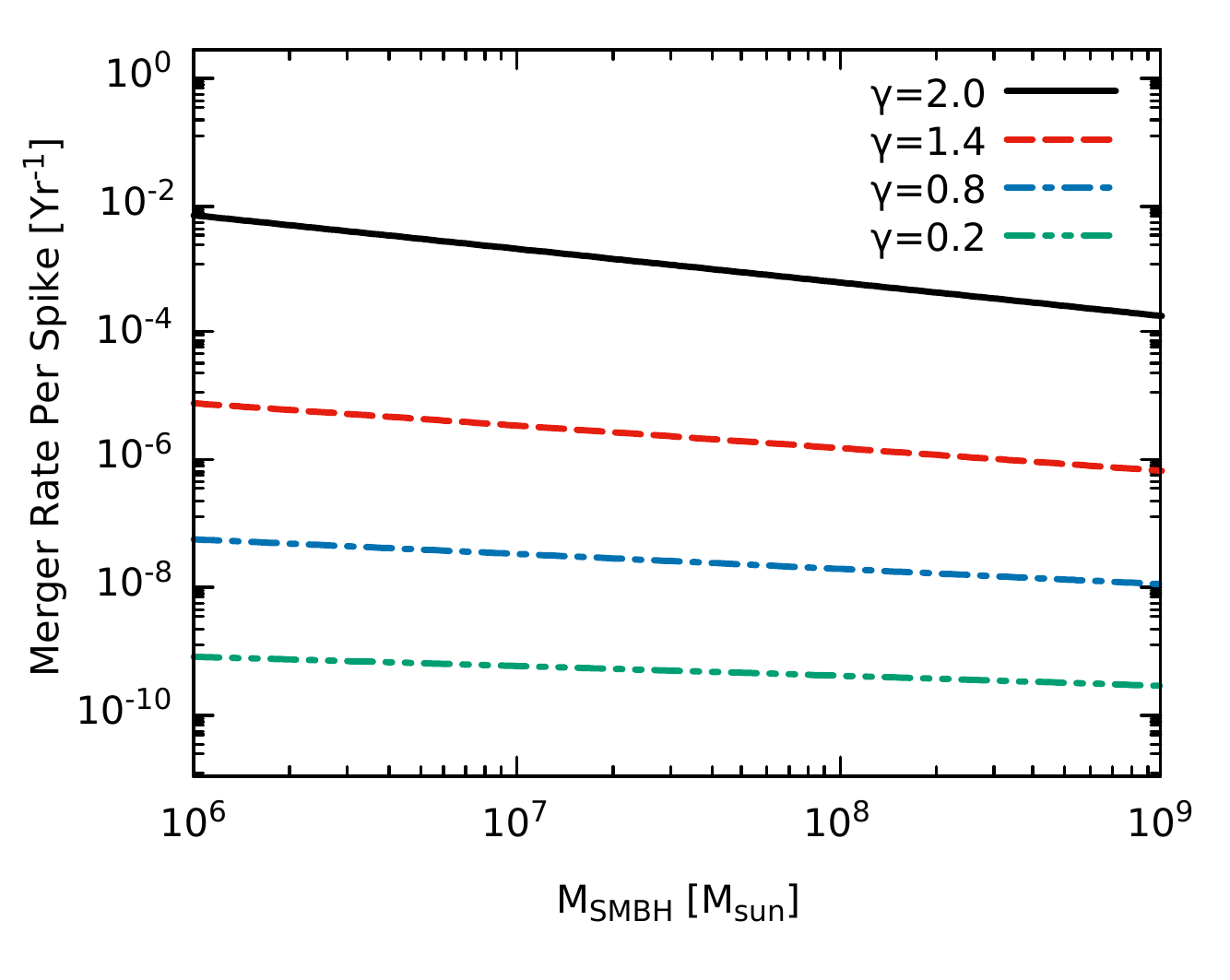}{0.45\textwidth}{(a) PBH-PBH -- Spherical Model}
\fig{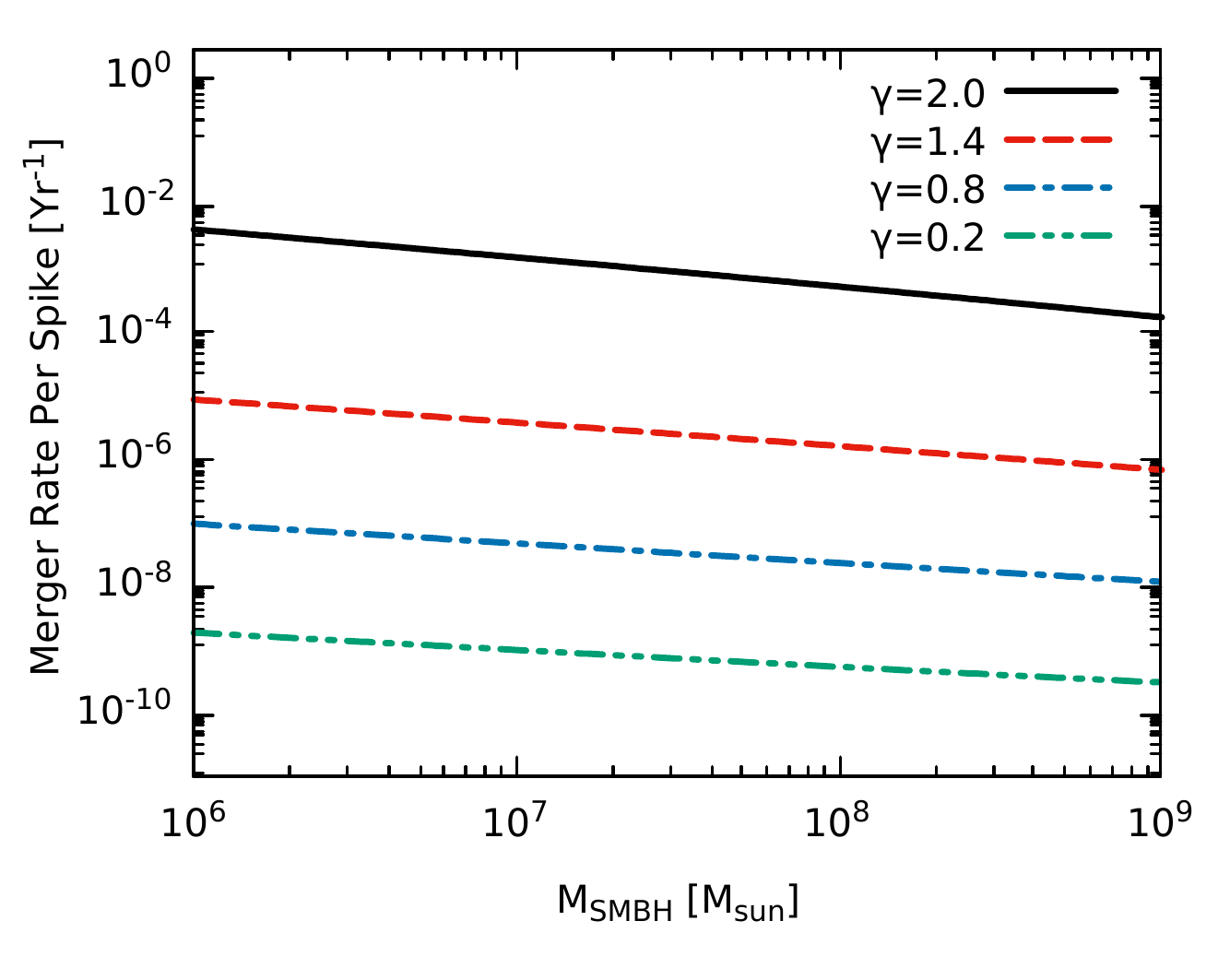}{0.45\textwidth}{(b) PBH-PBH -- Ellipsoidal Model}
}
\gridline{\fig{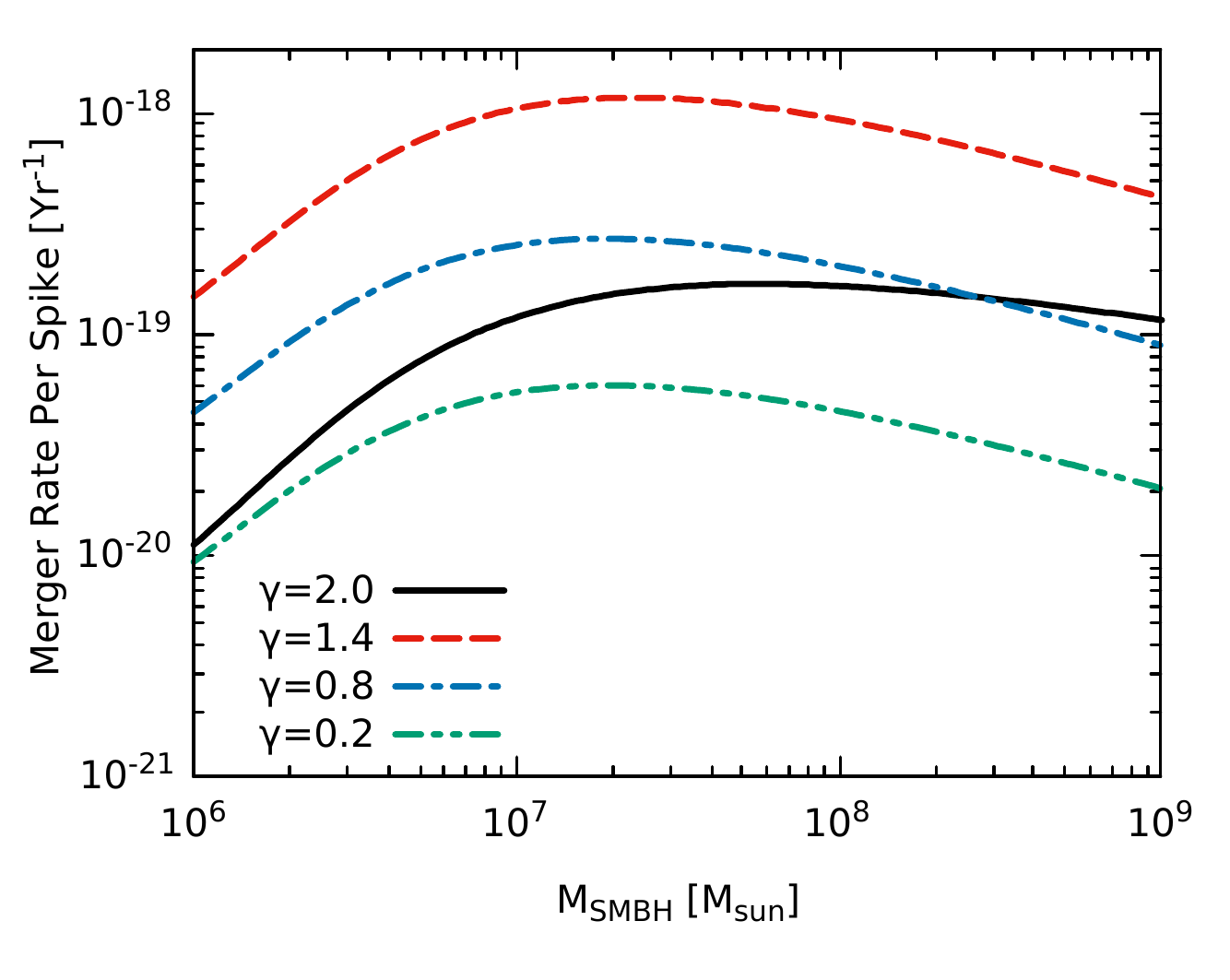}{0.45\textwidth}{(c) PBH-NS -- Spherical Model}
\fig{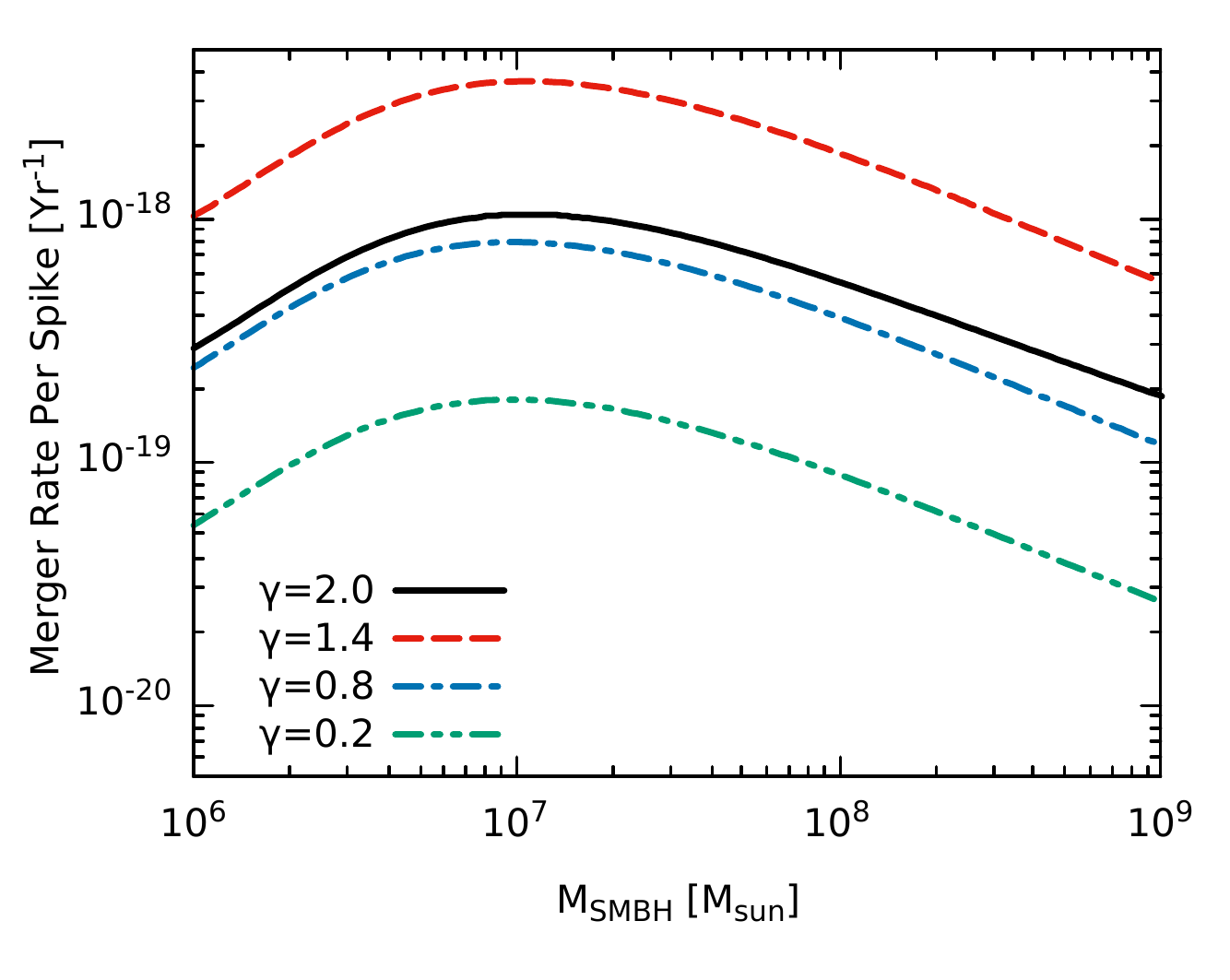}{0.45\textwidth}{(d) PBH-NS -- Ellipsoidal Model}
}
\caption{The merger rate of compact binaries in a single dark-matter spike as a function of SMBH mass for different values of $\gamma$. The top panels demonstrate this relation for PBH-PBH events in spherical- and ellipsoidal-collapse dark matter halo models, while the bottom panels display the corresponding results for PBH-NS events.}
\label{fig:2}
\end{figure*}

Therefore, the merger rate of compact binaries within each region of the dark-matter spike is determined as follows
\begin{equation}\label{merger_per_s}
N_{\rm sp} = 4\pi\int_{4r_{\rm s}}^{r_{\rm sp}} \mathcal{T}(\rho, m_1, m_2)\langle\xi v_{\rm rel}\rangle\, r^{2}dr,
\end{equation}
where for the PBH-PBH events:
\begin{equation} \label{condition1}
\mathcal{T}= \left\{\frac{1}{2}\frac{\left[f_{\rm PBH}\,\rho_{\rm sp}(r)\right]^{2}}{m_{1}m_{2}}\right\},
\end{equation}
and for the PBH-NS events:
\begin{equation}  \label{condition2}
\mathcal{T}= \left(\frac{f_{\rm PBH}\,\rho_{\rm sp}(r)}{m_{1}}\right)\left(\frac{\rho_{\rm NS}(r)}{m_{2}}\right).
\end{equation}
In the above relation, $0<f_{\rm PBH}\leq 1$ represents the fraction of PBHs that specifies their contribution to dark matter, and the angle bracket denotes an average over the relative velocity distribution at the vicinity of the central SMBH. Furthermore, $\rho_{\rm NS}(r)$ is the NS density profile that we define through the spherically symmetric form:
\begin{equation}
\rho_{\rm NS} = \rho^{*}_{\rm NS}\exp\left(-\frac{r}{r^{*}_{\rm NS}}\right),
\end{equation}
where $r^*_{\rm NS}$ and $\rho^*_{\rm NS}$ are characteristic radius and density of NSs, respectively, which need to be determined. For the characteristic radius of NSs, an approximative value has been proposed as $r^*_{\rm NS}\simeq 0.1\,r_{\rm s}$ \citep{2022ApJ...931....2S}, which we use in our calculations.

\begin{figure*}[ht!]
\gridline{\fig{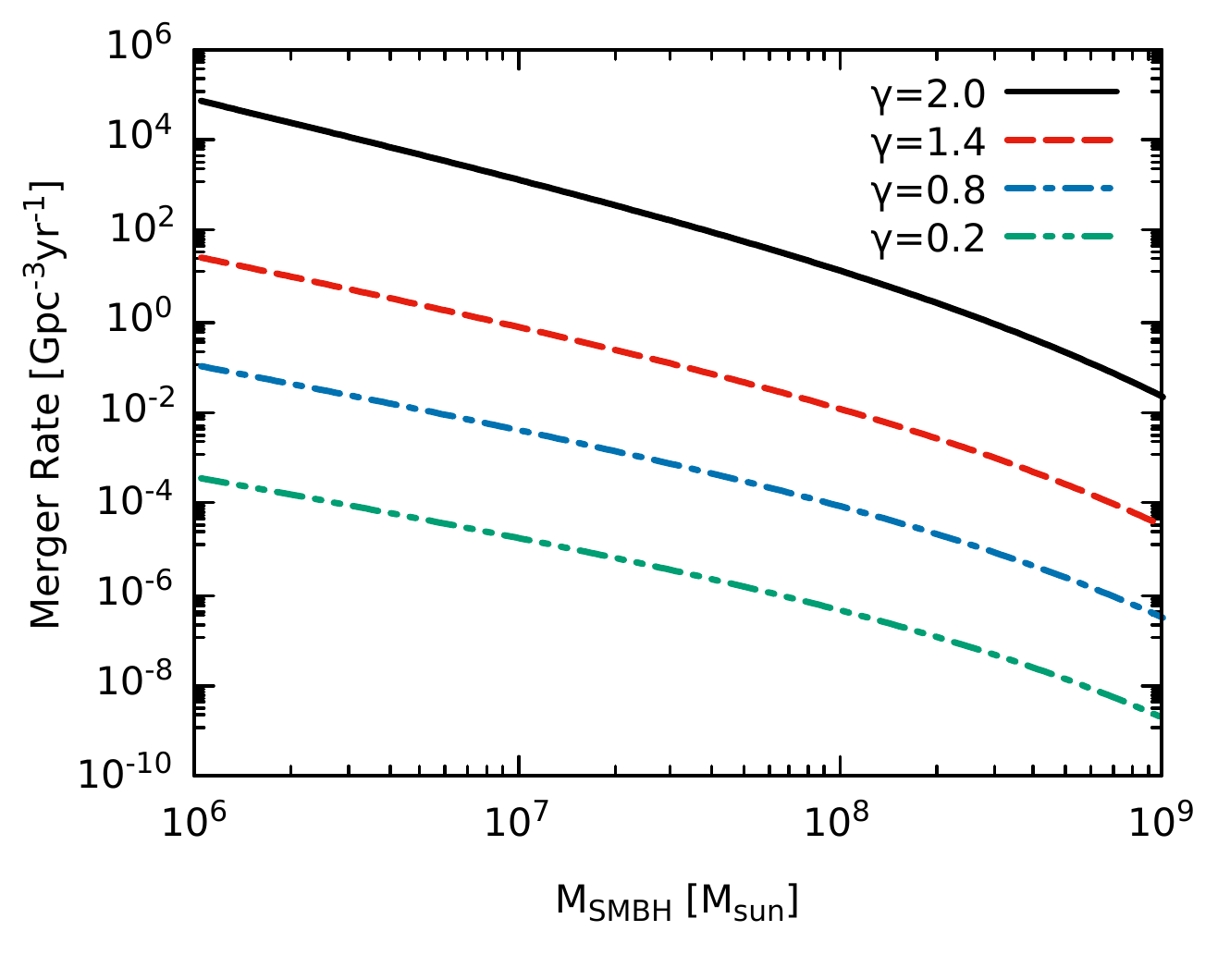}{0.33\textwidth}{(a) PBH-PBH -- Spherical -- Shankar M.F.}
\fig{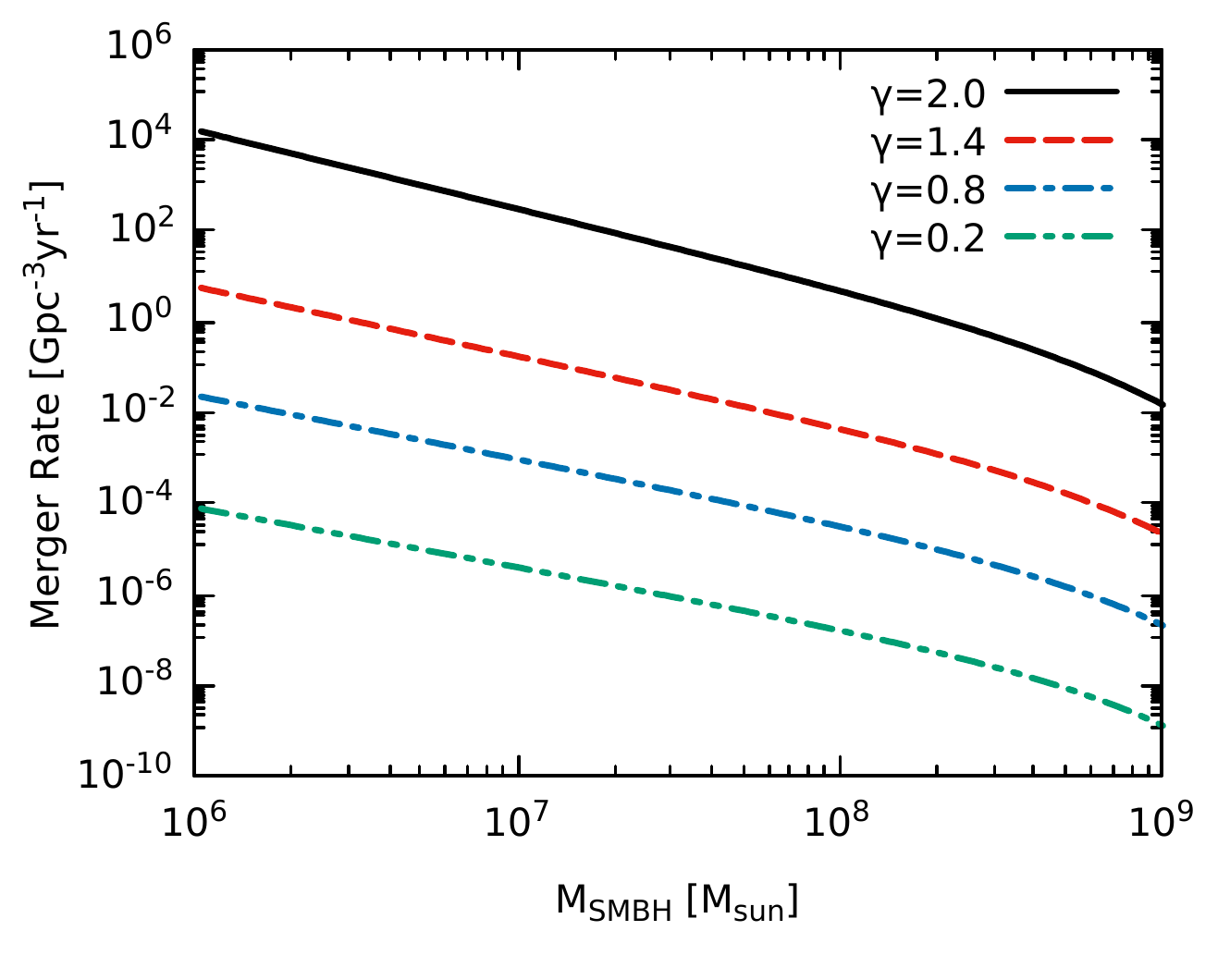}{0.33\textwidth}{(b) PBH-PBH -- Spherical -- Vika M.F.}
\fig{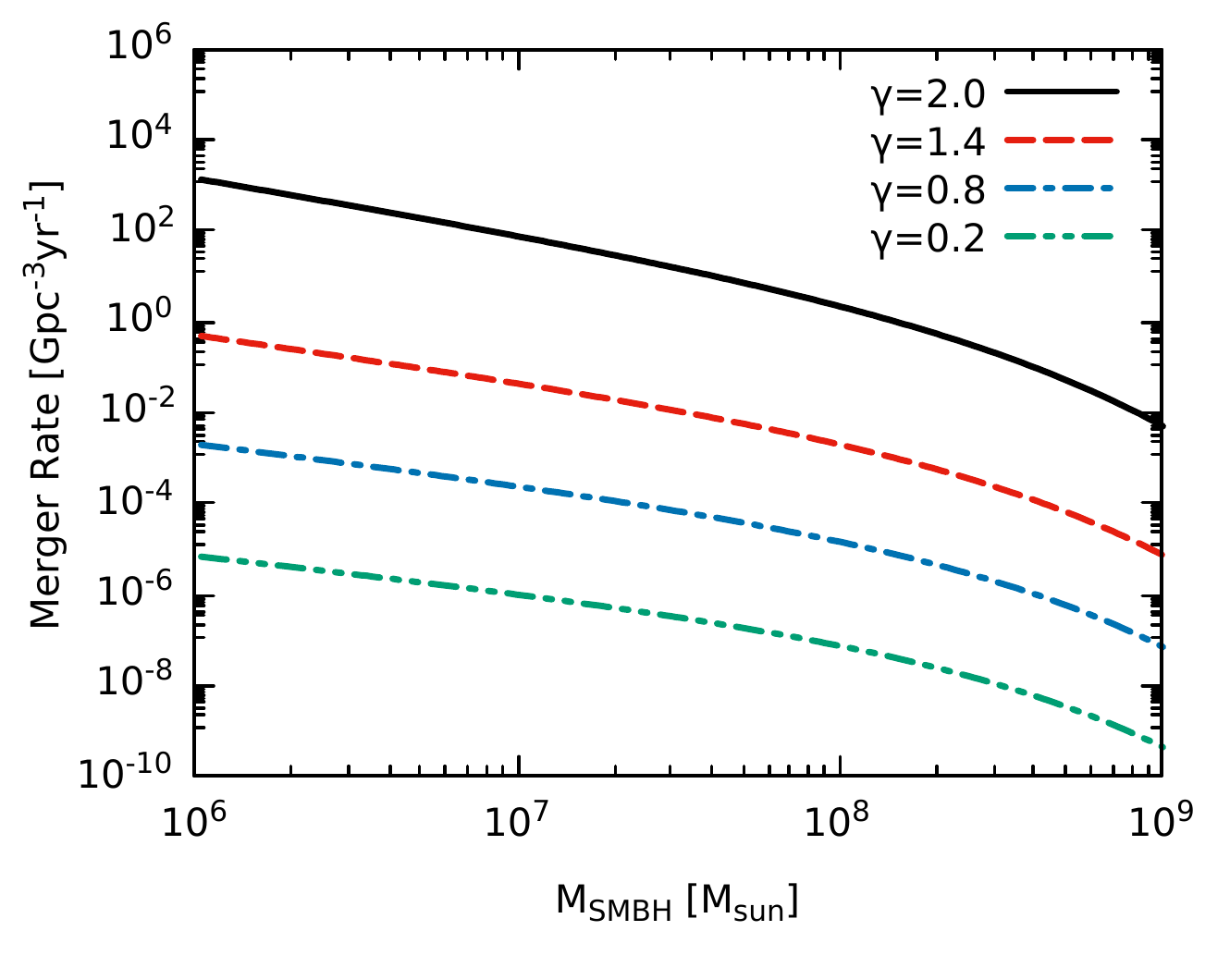}{0.33\textwidth}{(c) PBH-PBH -- Spherical -- Benson M.F.}
}
\gridline{\fig{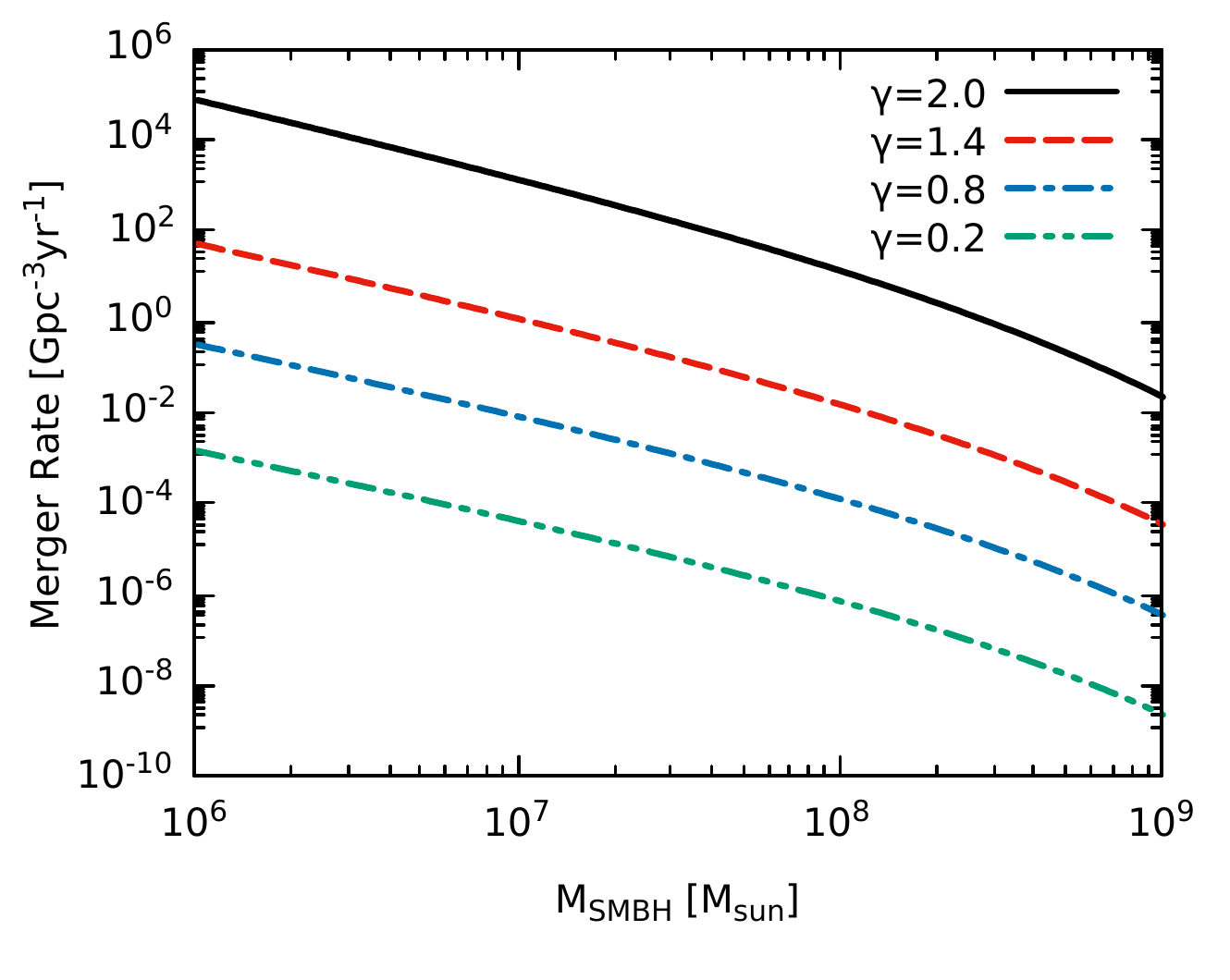}{0.33\textwidth}{(d) PBH-PBH -- Ellipsoidal -- Shankar M.F.}
\fig{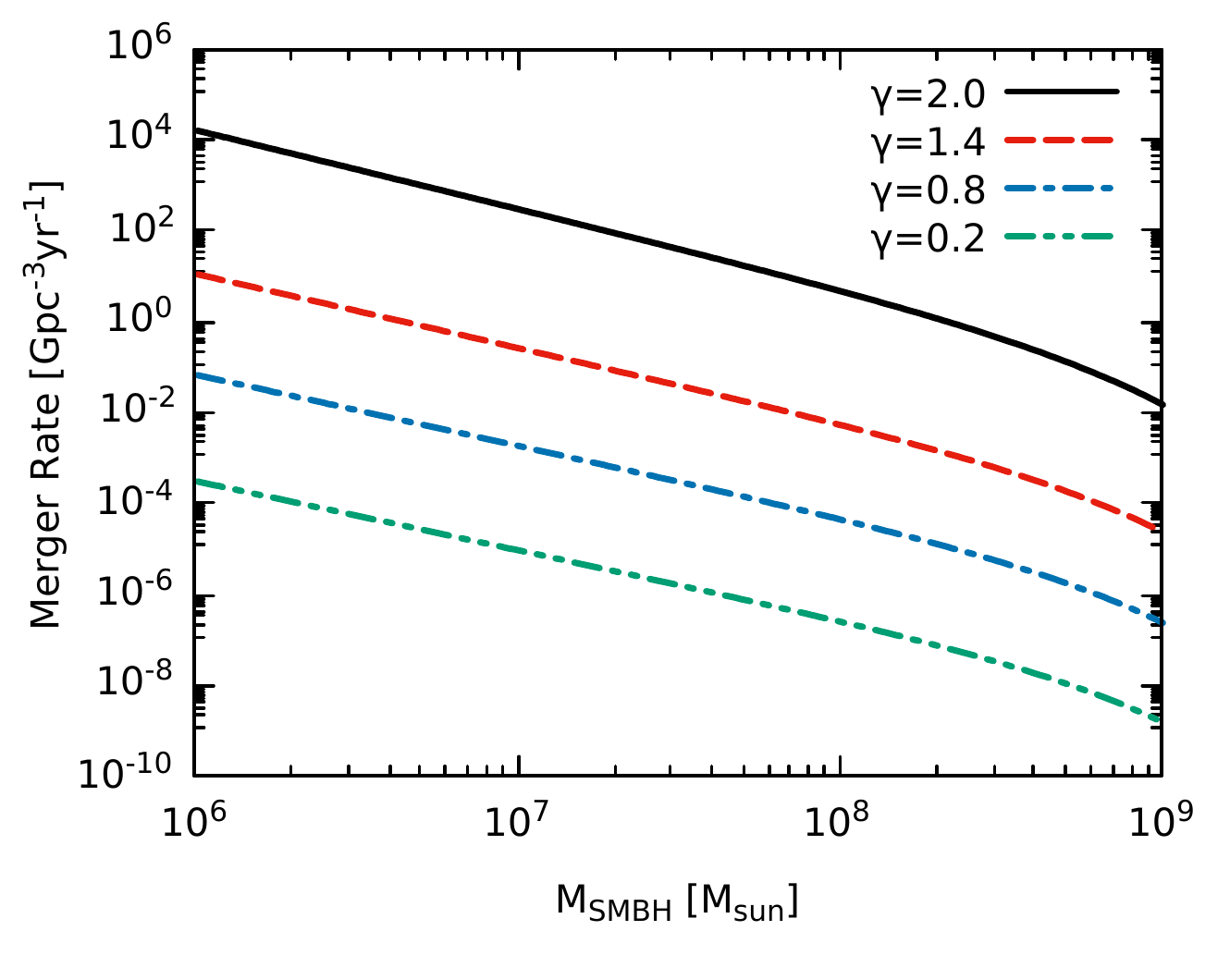}{0.33\textwidth}{(e) PBH-PBH -- Ellipsoidal -- Vika M.F.}
\fig{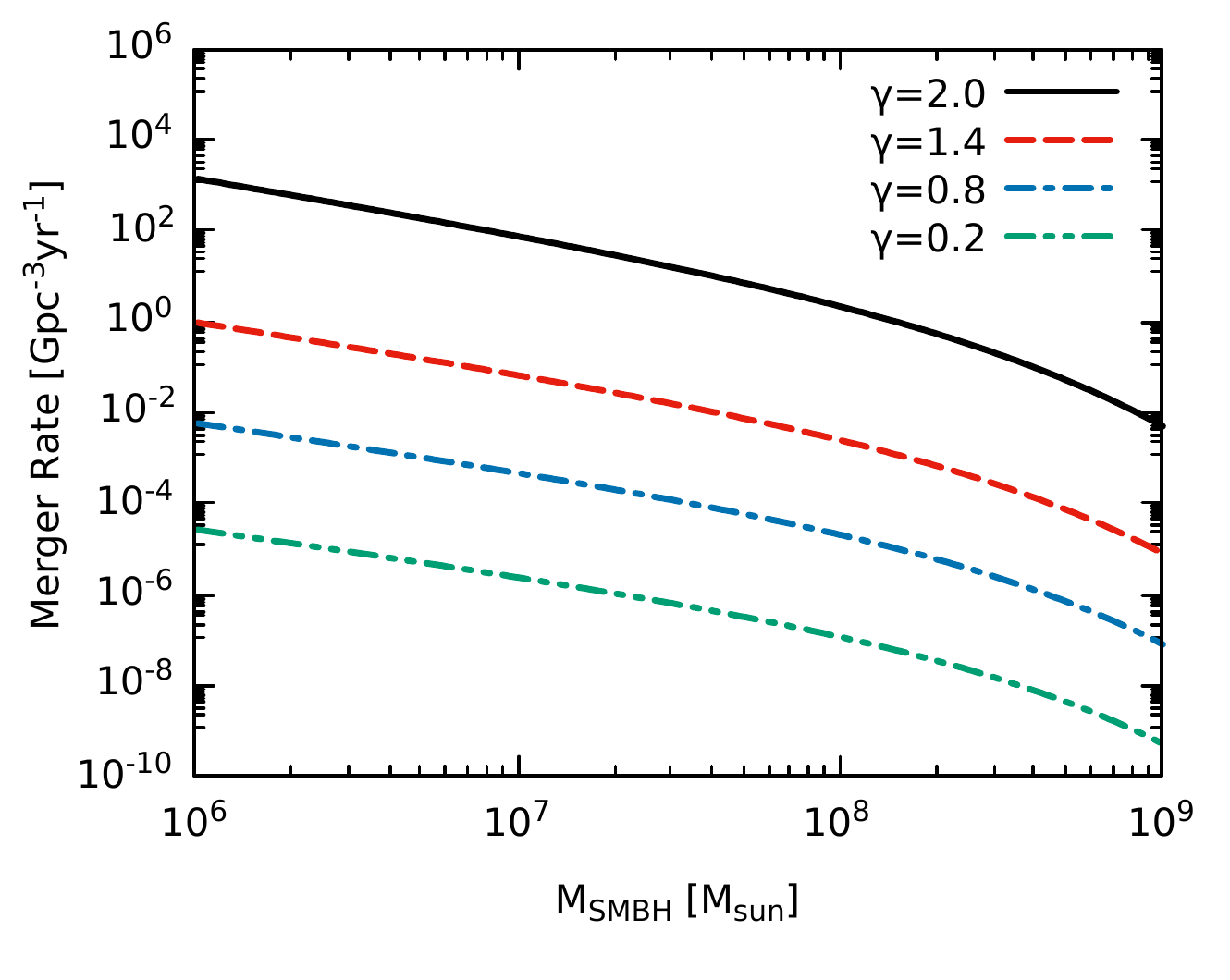}{0.33\textwidth}{(f) PBH-PBH -- Ellipsoidal -- Benson M.F.}
}
\caption{The merger rate of PBH-PBH binaries per unit time and volume as a function of SMBH mass for different values of $\gamma$. The top panels show this relation for spherical-collapse dark matter halo models while considering three different SMBH mass functions, whereas the bottom panels exhibit the corresponding results for ellipsoidal-collapse dark matter halo models.}
\label{fig:3}
\end{figure*}

Furthermore, the characteristic density of NSs must be obtained by normalizing the distribution of NSs to their estimated population in an arbitrary galaxy. To accomplish this, we utilize the time-independent form of the initial Salpeter stellar mass function, which is in the form $\chi(m_*) \approx m_*^{-2.35}$. Our main assumption is based on the fact that the entire population of stars in the mass range of $8M_{\odot}\mbox{-}20M_{\odot}$ will eventually yield a supernova explosion, and their outcome will be an NS. Consequently, the number of NSs in a single galaxy with stellar mass $M_{*}$ is given by
\begin{equation}
n_{\rm NS} = M_*\int_{m^{\rm min}_*}^{m^{\rm max}_*} \chi(m_*)dm_*,
\end{equation}
where $\chi(m_*)m_*$ is normalized to unity. It should be noted that to characterize the galactic stellar mass $M_*$, the stellar mass–halo mass relation $M_*(M_{\rm halo})$ must be determined. For this purpose, the stellar mass–halo mass relation obtained in \cite{2013ApJ...770...57B} can be used, the basic assumption of which is the presence of the maximum number of NSs at the center of the galactic halo. One must note that in the present analysis, for the relative velocity near the central SMBH, we use the circular velocity $v(r)=\sqrt{GM_{\rm SMBH}/r}$ at each radius bounded by the dark-matter spike. This is a reasonable choice since the total mass enclosed by the region of dark-matter spike is negligible versus the mass of the central SMBH. We consider the mass of involving PBHs in PBH-PBH events as $M_{\rm PBH}=30\,M_{\odot}$ and fix the masses of PBHs and NSs participating in PBH-NS events as $M_{\rm PBH}=5\,M_{\odot}$ and $M_{\rm NS}=1.4\,M_{\odot}$. Also, in the present analysis, we consider the contribution of PBHs in dark matter to be $f_{\rm PBH}=1$. It is obvious from Eqs. (\ref{merger_per_s}), (\ref{condition1}) and (\ref{condition2}) that the merger rate of PBH-PBH binaries is straightly proportional to the $f_{\rm PBH}^{2}$, while it changes directly with $f_{\rm PBH}$ for the PBH-NS events.

In Fig.\,\ref{fig:2}, we have plotted the merger rate of compact binaries within a single spike as a function of SMBH mass for several values of power-law index $\gamma$ while accounting for dark matter halo models with spherical and ellipsoidal collapses. As can be seen from the figure, the merger rate of PBH-PBH binaries changes inversely with the mass of the SMBH for both spherical- and ellipsoidal-collapse dark matter halo models. It is also evident that the merger rate of PBH-PBH binaries is directly proportional to the value of the power-law index $\gamma$. However, comparing PBH-PBH events for spherical- and ellipsoidal-collapse dark matter halo models, it can be inferred that the merger rate per spike for ellipsoidal-collapse dark matter halo models is higher than the corresponding one derived from spherical-collapse dark matter halo models.

\begin{figure*}[ht!]
\gridline{\fig{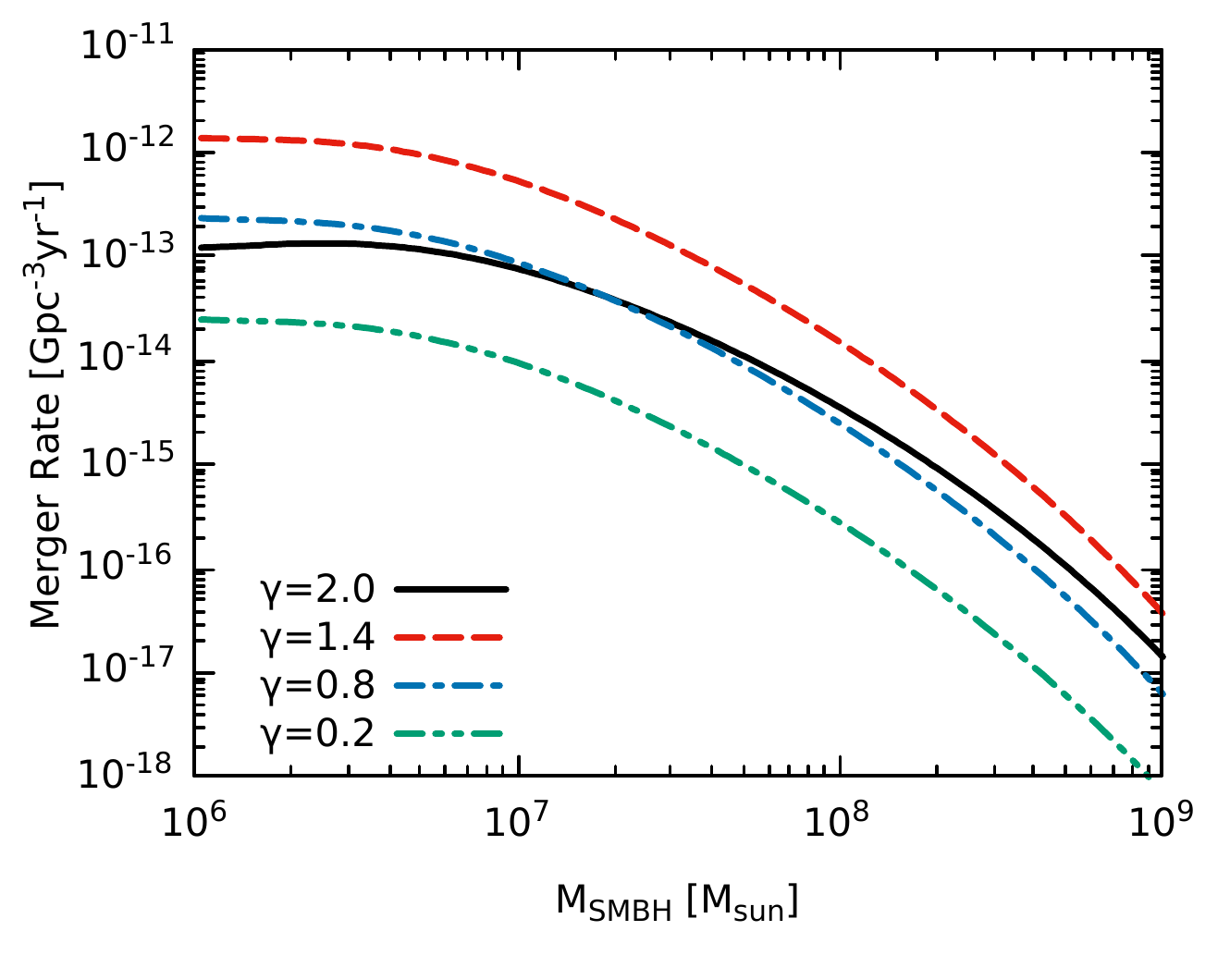}{0.33\textwidth}{(a) PBH-NS -- Spherical -- Shankar M.F.}
\fig{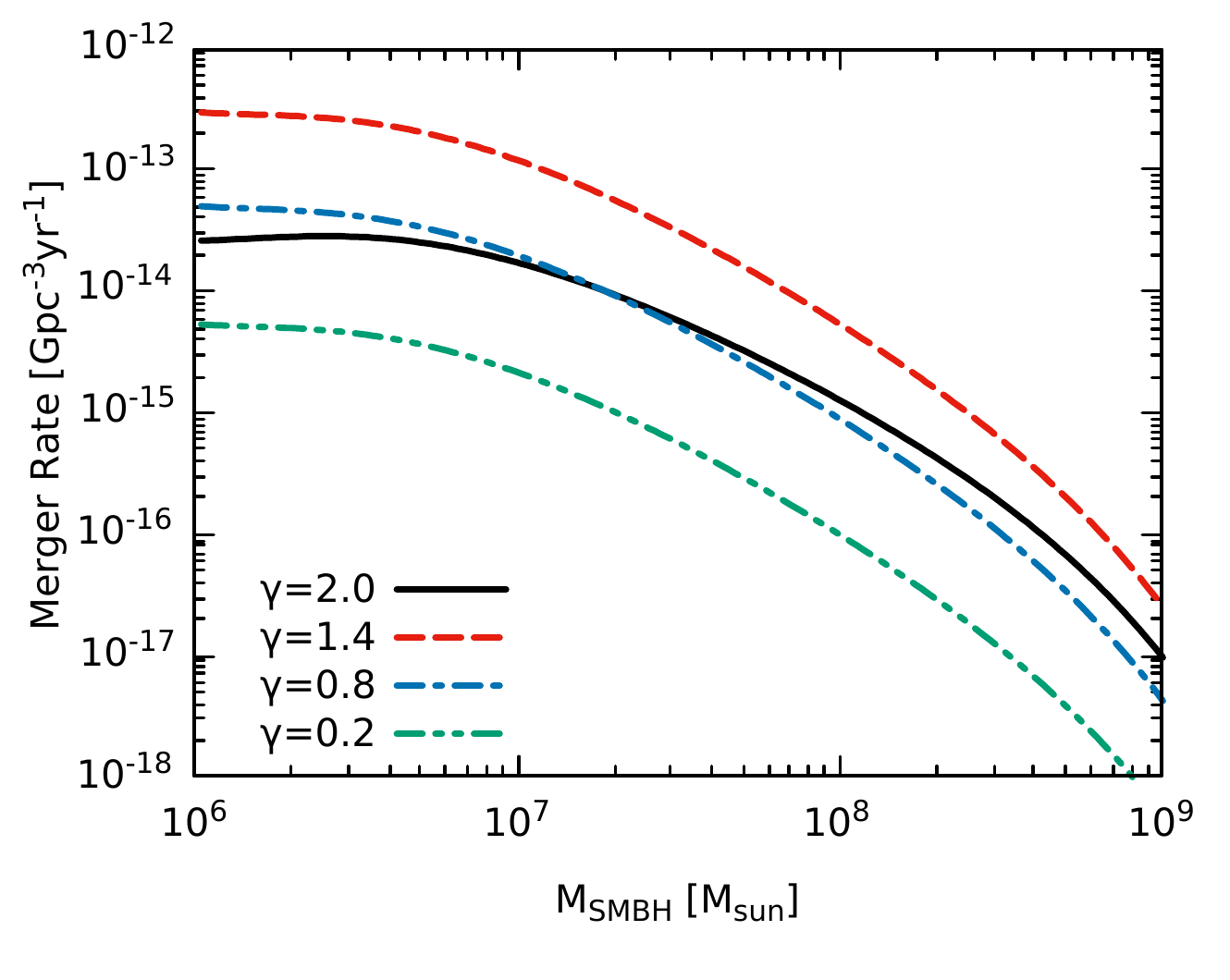}{0.33\textwidth}{(b) PBH-NS -- Spherical -- Vika M.F.}
\fig{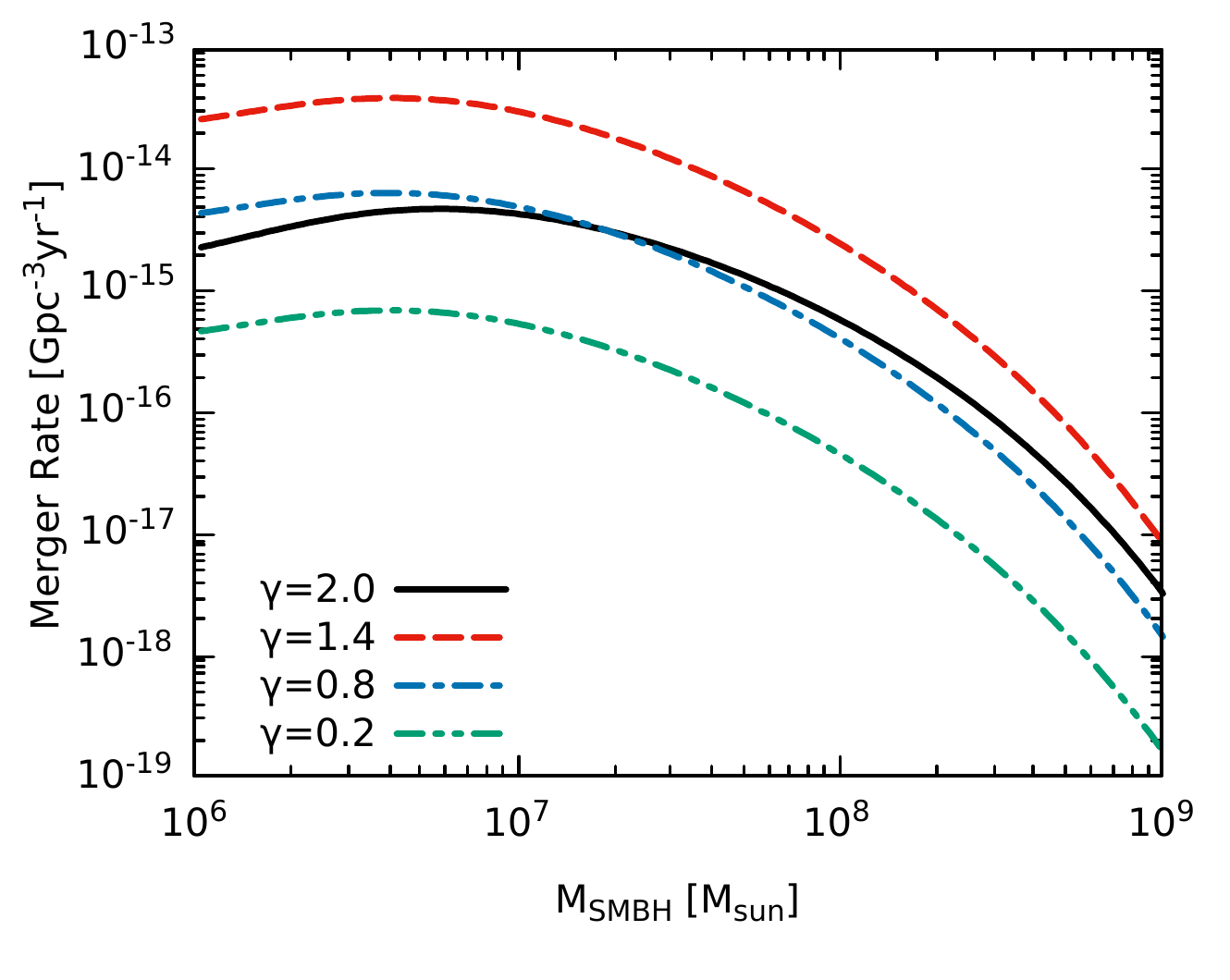}{0.33\textwidth}{(c) PBH-NS -- Spherical -- Benson M.F.}
}
\gridline{\fig{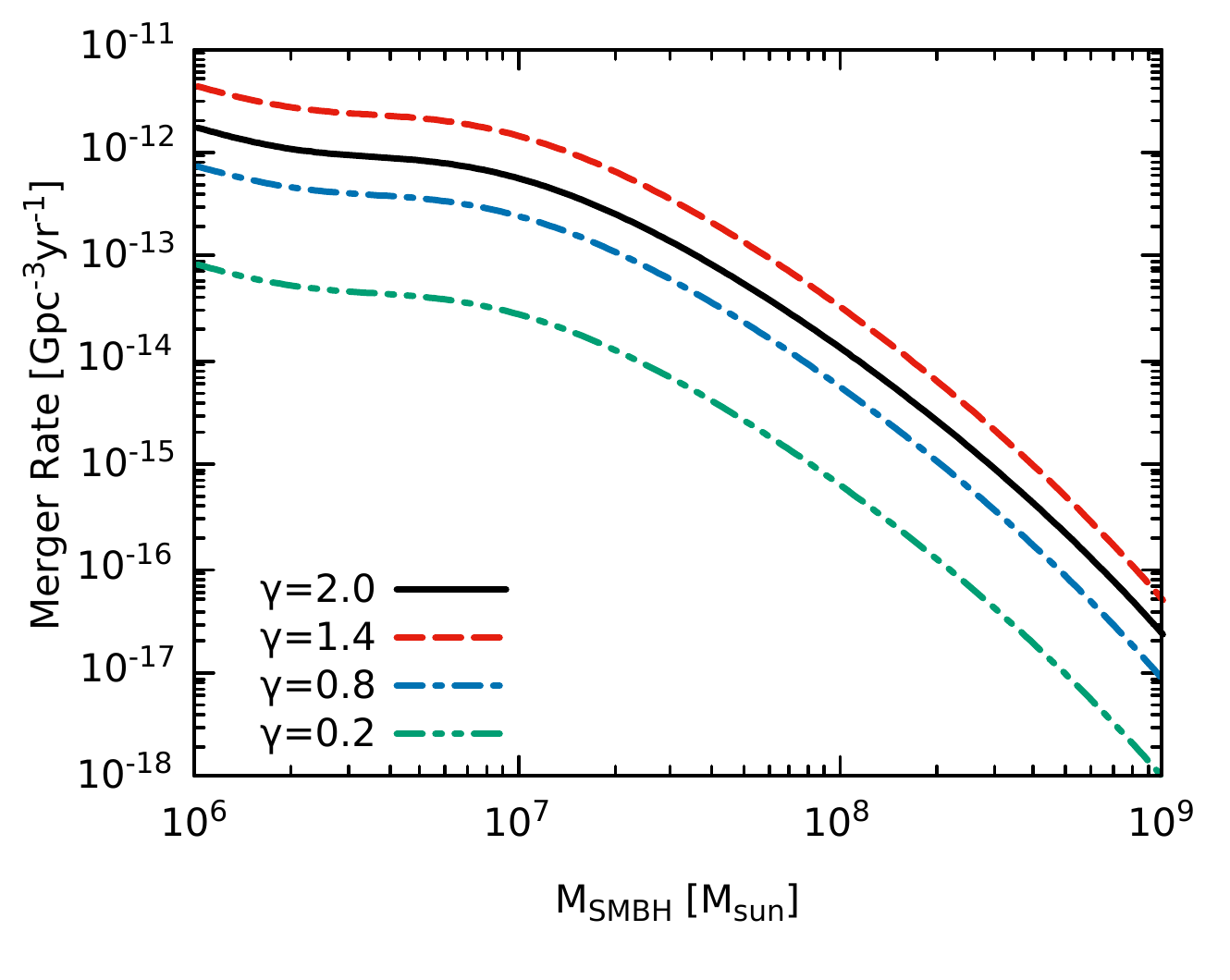}{0.33\textwidth}{(d) PBH-NS -- Ellipsoidal -- Shankar M.F.}
\fig{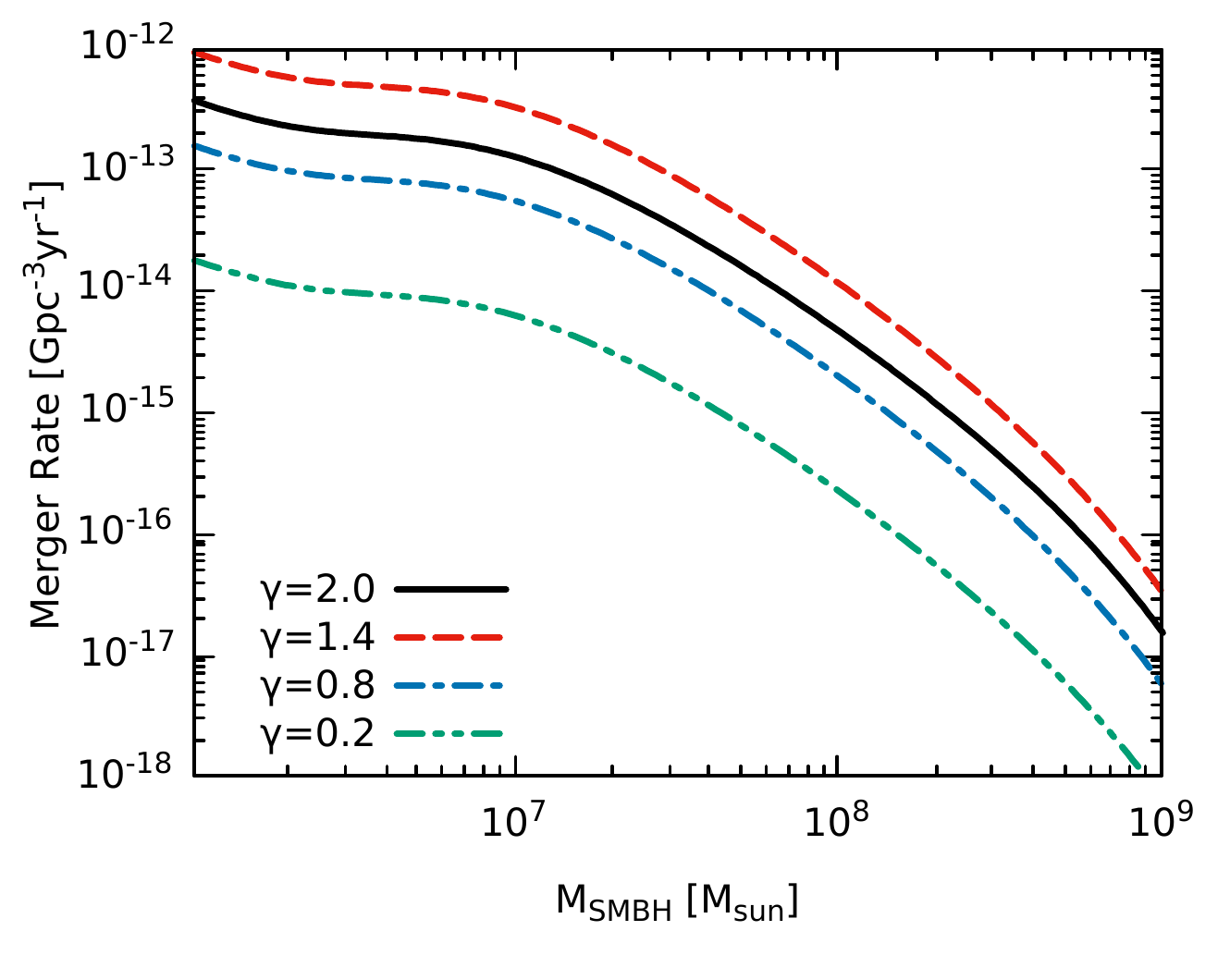}{0.33\textwidth}{(e) PBH-NS -- Ellipsoidal -- Vika M.F.}
\fig{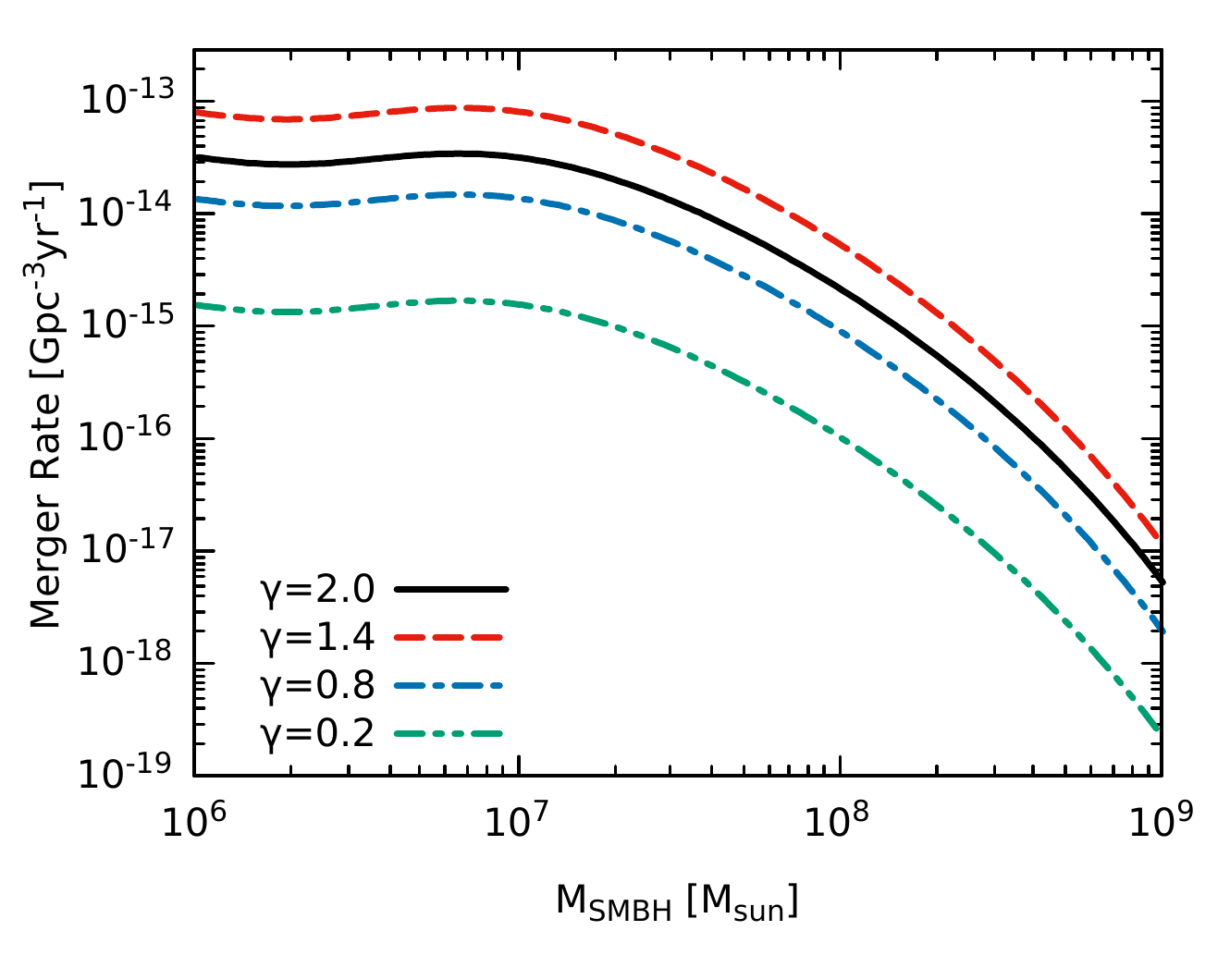}{0.33\textwidth}{(f) PBH-NS -- Ellipsoidal -- Benson M.F.}
}
\caption{The merger rate of PBH-NS binaries per unit time and volume as a function of SMBH mass for different values of $\gamma$. The top panels indicate this relation for spherical-collapse dark matter halo models while accounting for three different SMBH mass functions, whereas the bottom panels show the corresponding results for ellipsoidal-collapse dark matter halo models.}
\label{fig:4}
\end{figure*}

However, for the merger rate of PBH-NS binaries, the situation is slightly different. Interestingly, unlike the previous case, the merger rate of PBH-NS binaries per spike reaches the maximum value for the SMBH mass of $M_{\rm SMBH}\simeq 10^{7}M_{\odot}$ for both spherical- and ellipsoidal-collapse dark matter halo models. Also, it can be deduced from the calculation that the merger rate of PBH-NS binaries per spike for a power-law index of $\gamma=1.7$ has a maximum value that is completely different from the relevant results of PBH-PBH events. In other words, the merger rate of PBH-NS binaries increases monotonically up to $\gamma=1.7$ and decreases for the larger values. In the following, we will discuss this in more detail. In addition, it can be inferred from the results that the merger rate of PBH-NS binaries for ellipsoidal-collapse dark matter halo models is much higher than the corresponding one derived from spherical-collapse dark matter halo models.

On the other hand, the cumulative merger rate of compact binaries is considered the main quantity to be recorded and processed through the LIGO-Virgo detectors. Therefore, the overall merger rate of compact binaries per unit volume and per unit time needs to be specified. To perform this task, one has to convolve the mas function of SMBH, $\phi(M_{\rm SMBH})$, with the merger rate of compact binaries per spike, $N_{\rm sp}(M_{\rm SMBH})$:
\begin{equation}
\mathcal{R} = \int_{M_{\rm min}}^{M_{\rm max}} N_{\rm sp}(M_{\rm SMBH}) \phi(M_{\rm SMBH}) dM_{\rm SMBH}.
\end{equation}

According to Eqs.\,(\ref{massfunc1}), (\ref{massfunc2}) and (\ref{massfunc3}), the mentioned mass functions have a decreasing exponential term with respect to the mass of SMBHs. Therefore, it can be concluded that $M_{\rm max}$ does not have a significant effect on the final result. In contrast, the introduced mass functions indicate that the maximum abundance belongs to the smallest central black holes in the Universe. Consequently, $M_{\rm min}$ can have a significant contribution to the merger rate of compact binaries in dark-matter spikes.

In Fig.\,\ref{fig:3}, we have depicted the merger rate of PBH-PBH binaries per unit time and volume as a function of SMBH mass for several values of power-law index $\gamma$ while considering dark matter halo models with spherical and ellipsoidal collapses. We have provided the results for three mass functions \cite{2004MNRAS.354.1020S}, \cite{2007MNRAS.379..841B}, and \cite{2009MNRAS.400.1451V} to justify the possible uncertainties in the present analysis as much as possible. The merger rate of PBH-PBH binaries in both dark matter halo models with spherical and ellipsoidal collapses decreases monotonically with increasing the mass of SMBHs. Given the classification of the mass functions of SMBHs for their abundance, this result seems reasonable. As it is clear from the figures, the merger rate of PBH-PBH binaries for all three mass functions and both dark matter halo models reaches the maximum value as $M_{\rm SMBH}=10^{6}M_{\odot}$. Also, the direct proportionality of the merger rate of PBH-PBH binaries to the values of the power-law index is evident. Moreover, the merger rate of PBH-PBH binaries in ellipsoidal-collapse dark matter halo models is slightly higher than that obtained from spherical-collapse dark matter halo models. In the best case, which can be realized at the minimum value of the power-law index, e.g. $\gamma=0.05$, the amplification of the overall merger rate is $62\%$. Additionally, it can be concluded that the merger rate of PBH-PBH binaries yields the highest, middle, and lowest values while considering \cite{2004MNRAS.354.1020S}, \cite{2009MNRAS.400.1451V}, and \cite{2007MNRAS.379..841B} mass functions respectively.

\begin{figure*}[ht!]
\gridline{\fig{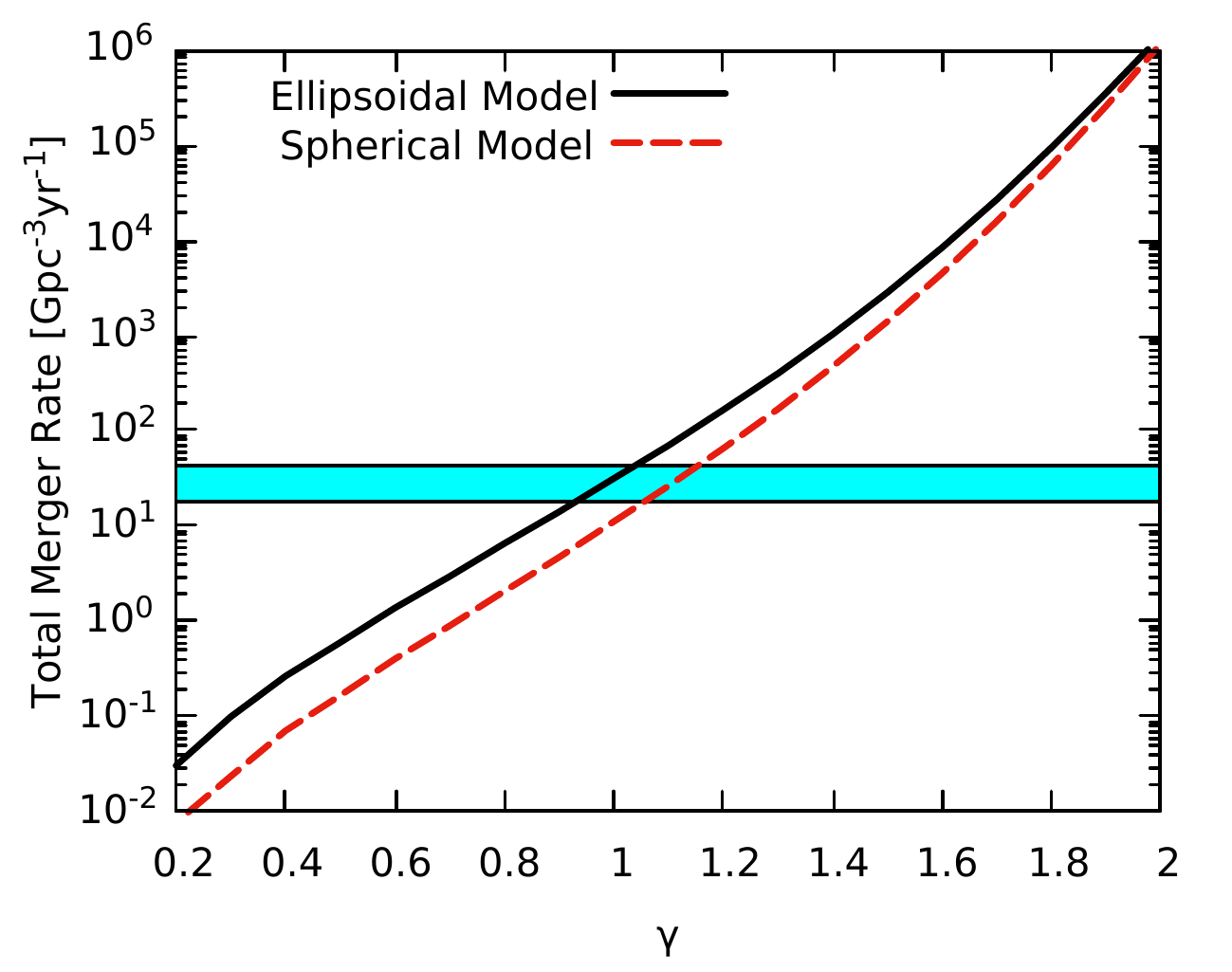}{0.33\textwidth}{(a) PBH-PBH -- Shankar M.F.}
\fig{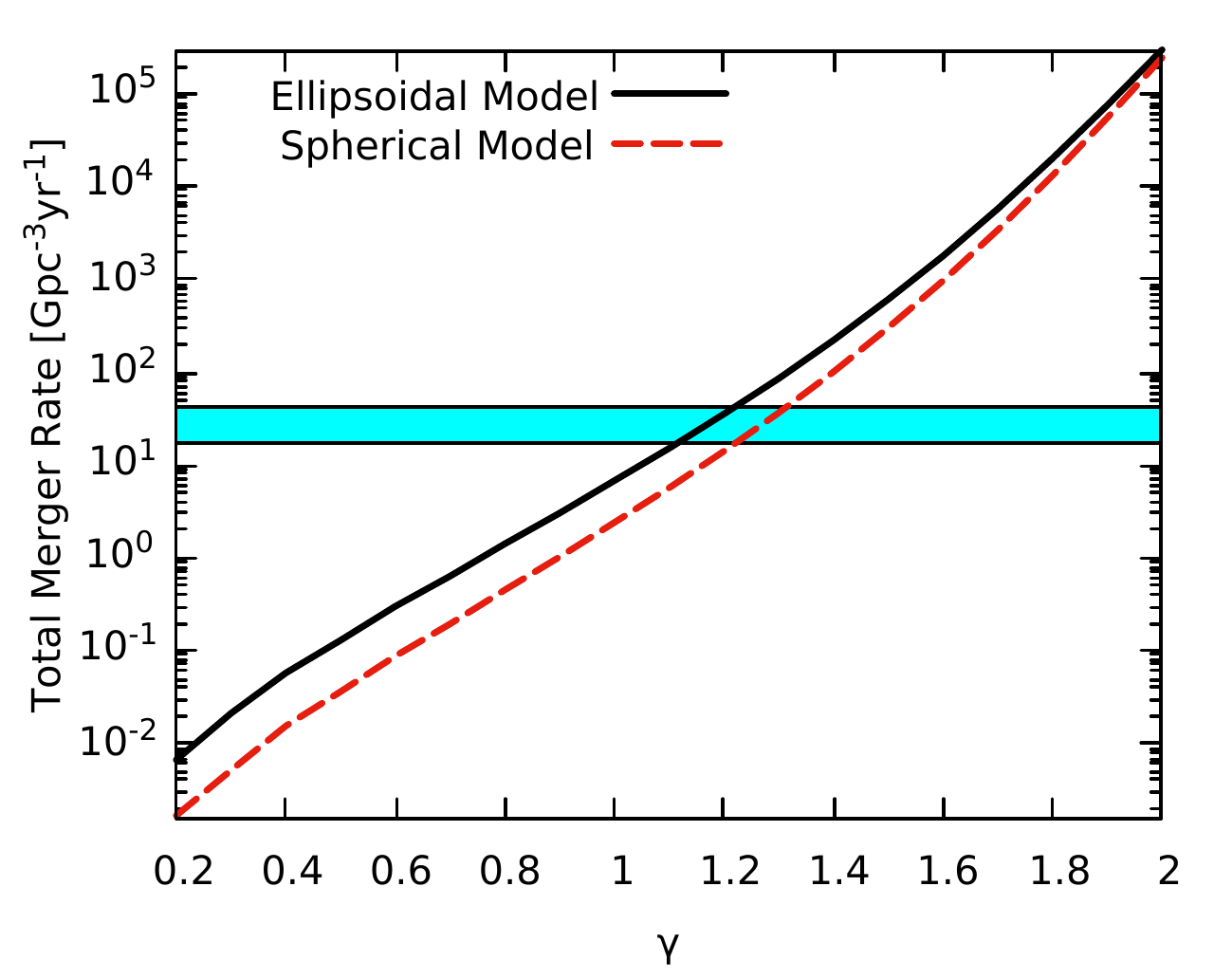}{0.33\textwidth}{(b) PBH-PBH -- Vika M.F.}
\fig{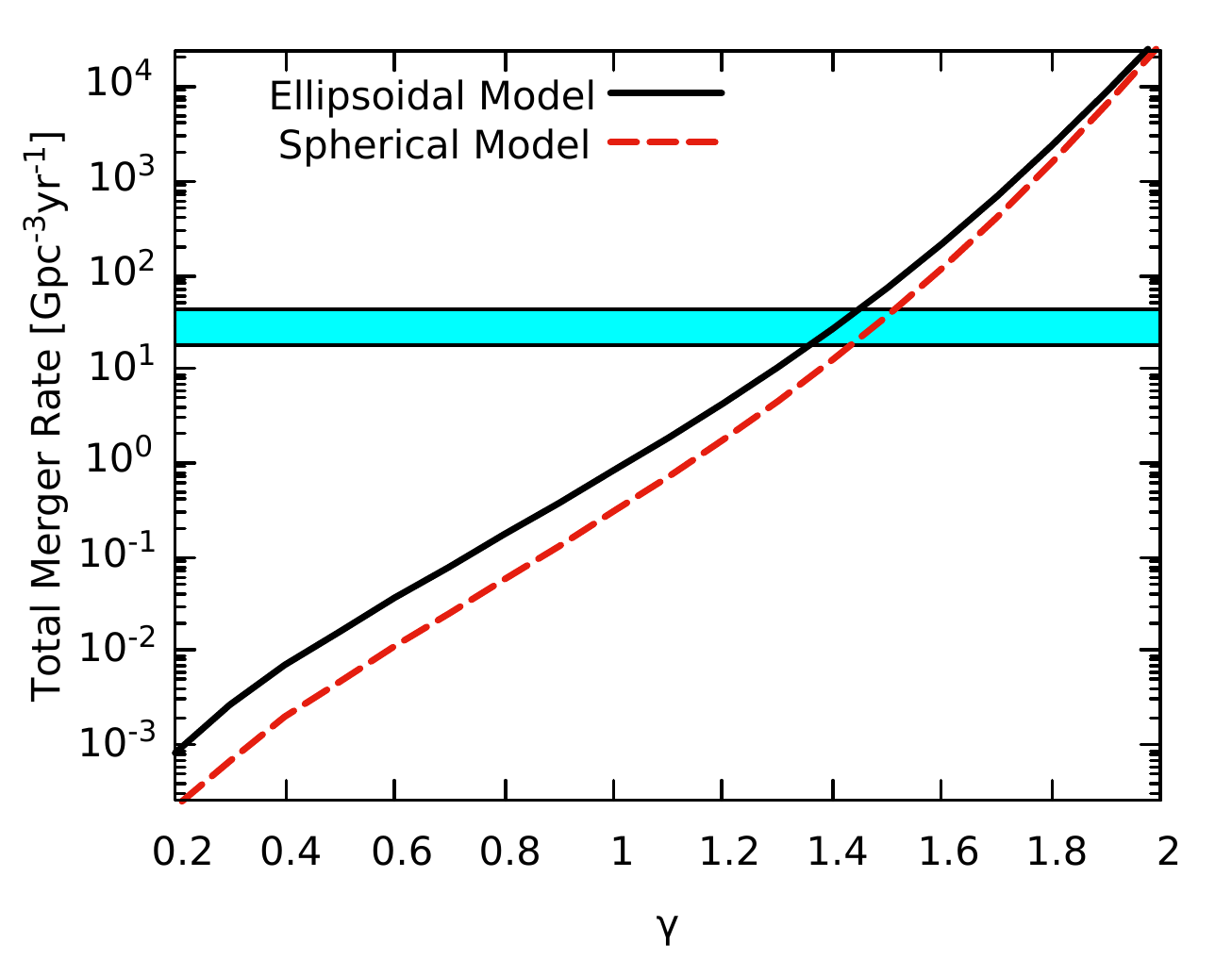}{0.33\textwidth}{(c) PBH-PBH -- Benson M.F.}
}
\gridline{\fig{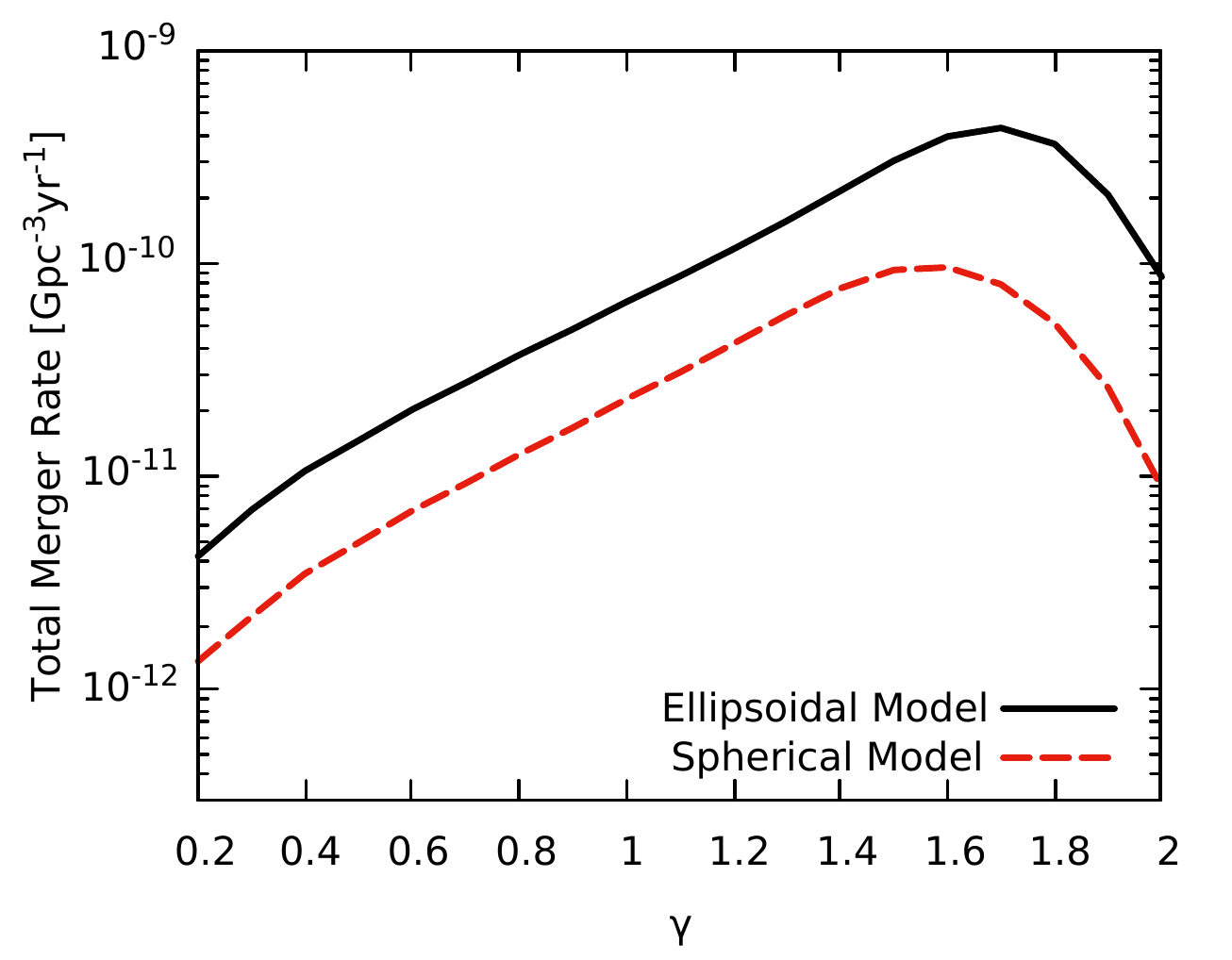}{0.33\textwidth}{(d) PBH-NS -- Shankar M.F.}
\fig{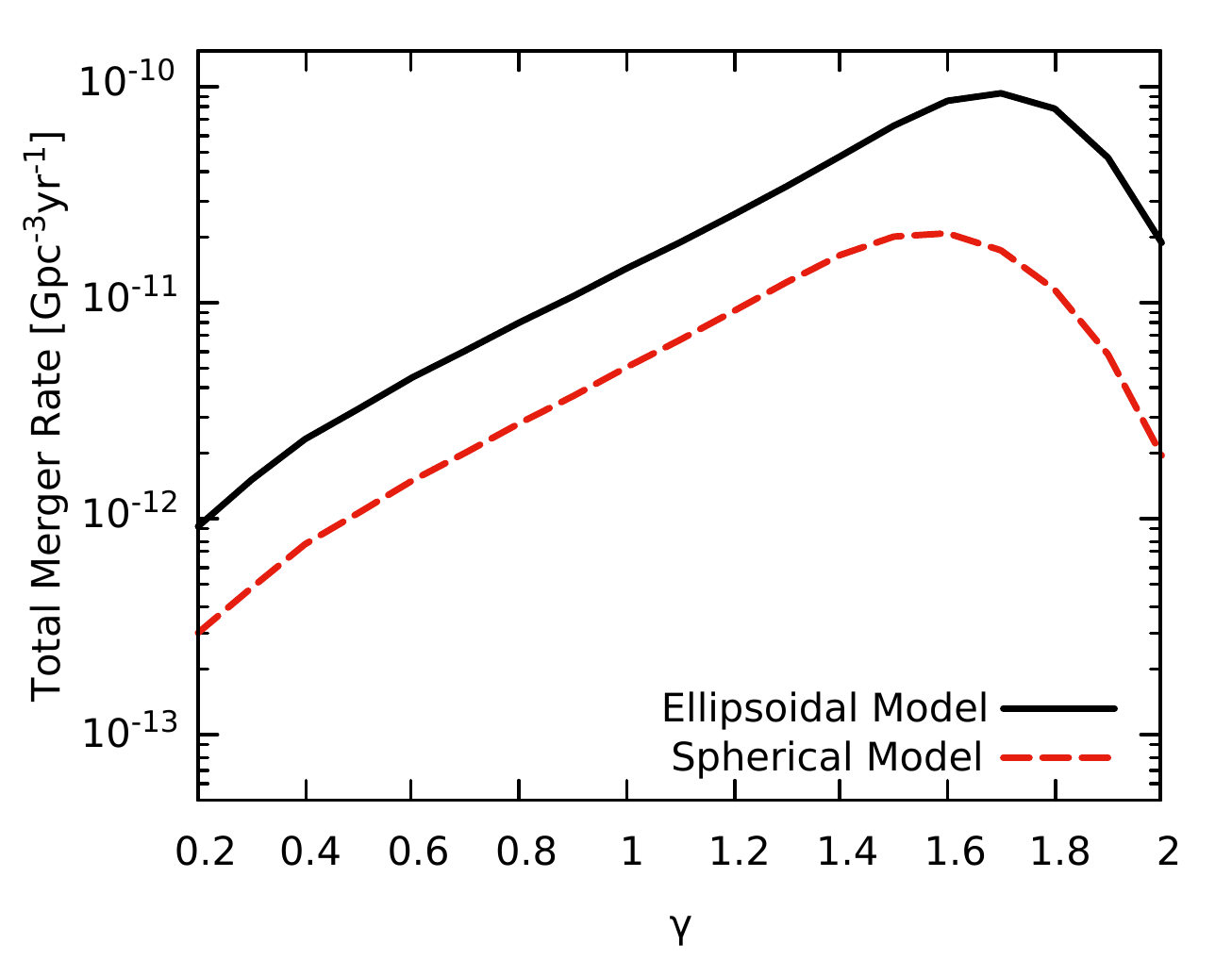}{0.33\textwidth}{(e) PBH-NS -- Vika M.F.}
\fig{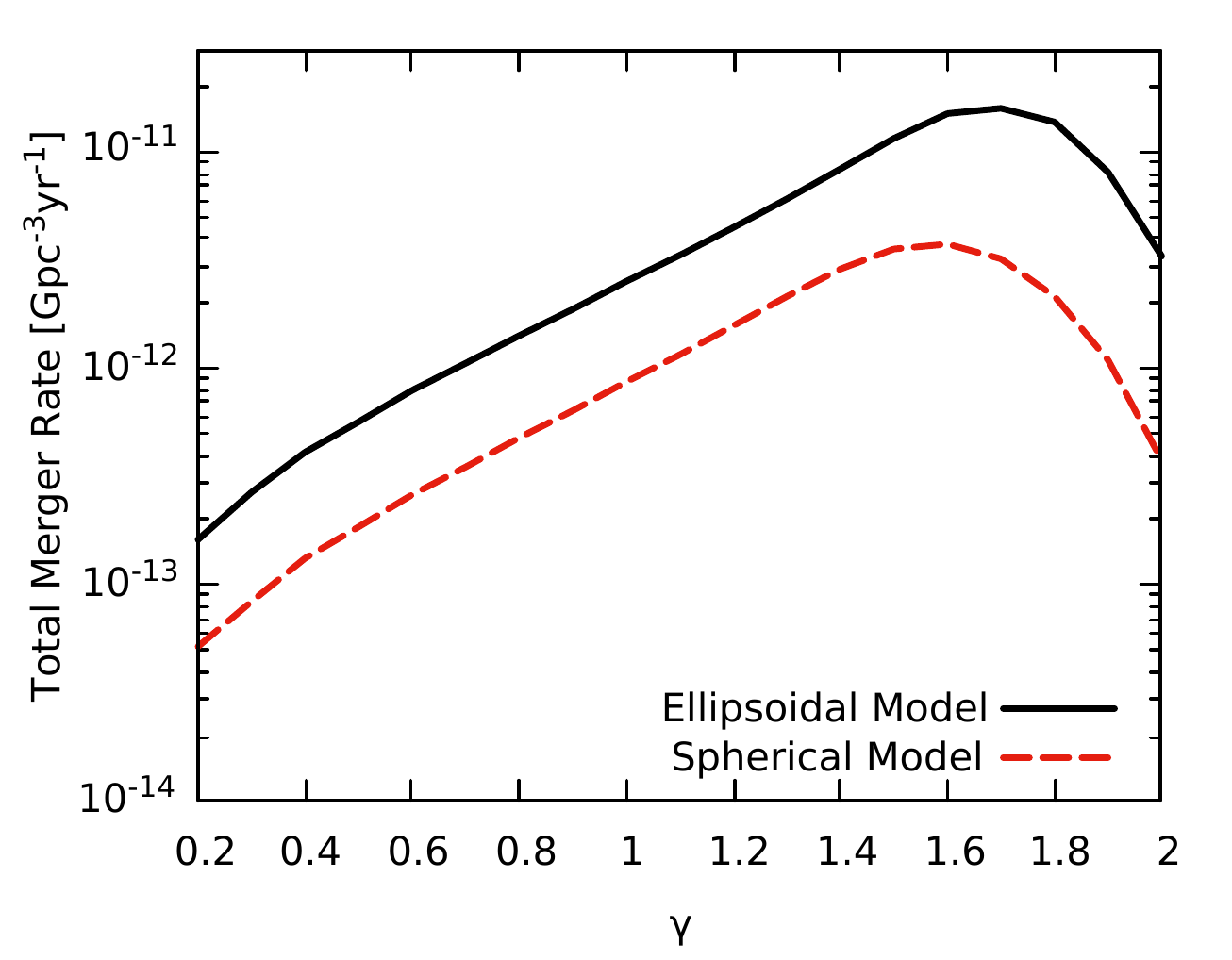}{0.33\textwidth}{(f) PBH-NS -- Benson M.F.}
}
\caption{The overall merger rate of compact binaries as a function of power-law index $\gamma$ for both spherical- and ellipsoidal-collapse dark matter halo models. The top panels demonstrate this relation for PBH-PBH events for three different SMBH mass functions, while the bottom panels show the corresponding results for PBH-NS events. The shaded cyan bands represent the BH-BH mergers estimated by the LIGO-Virgo detectors during the latest observing run, i.e., $(17.9\mbox{-}44)\,{\rm Gpc^{-3}Yr^{-1}}$.}
\label{fig:5}
\end{figure*}

Similarly to the merger rate per spike discussed earlier, Fig.\,\ref{fig:4} exhibits that the overall merger rate of PBH-NS binaries differs from that of PBH-PBH binaries. As a common point of whole applied dark matter models and mass functions, it is deduced that the merger rate of PBH-NS binaries has a maximum at $\gamma=1.7$ and experiences a decreasing behavior for values greater than that (see Fig.\,\ref{fig:5}). It is also evident that the merger rate of PBH-NS binaries for ellipsoidal-collapse dark matter halo models is in turn higher than that obtained for spherical-collapse dark matter halo models. Such a relative advantage could stem from the fact that, one can hope to improve the theoretical predictions of the merger rate of compact binaries in galactic halos by using more realistic dark matter halo models. Interestingly, for PBH-NS binaries, the inverse proportionality of the overall merger rate with the mass of SMBHs is not monotonic and has a plateau in some cases. If spherical-collapse dark matter halo models are reliable, the results obtained for \cite{2004MNRAS.354.1020S} and \cite{2009MNRAS.400.1451V} mass functions for $\gamma < 1.7$ decrease monotonically with the mass of PBHs, while for $\gamma > 1.7$, it has a plateau around $M_{\rm SMBH}=10^{7}M_{\odot}$. However, the result obtained from the \cite{2007MNRAS.379..841B} mass function has a plateau around the same SMBH mass for all values of $\gamma$. On the other hand, if the ellipsoidal-collapse dark matter halo models are plausible, the existence of the plateau is evident for all the mass functions used and for all values of the power-law index. These results suggest that dark-matter spikes structured around central SMBHs with mass $M_{\rm SMBH}=10^{7}M_{\odot}$ may contain a non-standard abundance of NSs that disturbs the monotonic dependence of the merger rate on the mass SMBHs.

In Fig.\,\ref{fig:5}, to quantitatively compare the results obtained from ellipsoidal-collapse dark matter halo models with those obtained from spherical-collapse dark matter halo models, we have displayed the merger rate of compact binaries in terms of power-law index $\gamma$ while taking into account \cite{2004MNRAS.354.1020S}, \cite{2009MNRAS.400.1451V}, and \cite{2007MNRAS.379..841B} mass functions. To assess the theoretical predictions via the experimental data, we have also included the relevant mergers estimated by the GW detectors in the case of PBH-PBH events, while such an action has not been taken in the case of PBH-NS events. This is because the prediction of the present analysis from the merger rate of PBH-PBH binaries is capable of justifying the data recorded by GW detectors, i.e., $(17.9\mbox{-}44)\,{\rm Gpc^{-3}Yr^{-1}}$ \citep{2021arXiv211103606T}, whereas the merger rate of PBH-NS binaries cannot in any way perform this task. Specifically, the estimate of the LIGO-Virgo detectors from the total merger rate of BH-NS binaries is presented as $(7.8\mbox{-}140)\,{\rm Gpc^{-3}Yr^{-1}}$ \citep{2021arXiv211103634T}, while the relevant prediction of both dark matter halo models in the current analysis is extremely far from this range.

As it is evident from the figures related to PBH-PBH events, and of course as mentioned earlier, the merger rate of PBH-PBH binaries while accounting for ellipsoidal-collapse dark matter halo models is higher than that extracted from spherical-collapse dark matter halo models. In addition, such enhancement in the merger rate is greater at lower power-law index values than that at higher ones. In this regard, the merger rate of PBH-PBH binaries in spherical-collapse dark matter halo models while considering \cite{2004MNRAS.354.1020S}, \cite{2009MNRAS.400.1451V}, and \cite{2007MNRAS.379..841B} mass functions will be consistent with the BH-BH mergers estimated by GW detectors if the value of power-law index lies in the interval $\gamma=(1.05\mbox{-}1.15)$, $\gamma=(1.10\mbox{-}1.20)$, and $\gamma=(1.40\mbox{-}1.50)$, respectively. However, the corresponding results obtained from ellipsoidal-collapse dark matter halo models can potentially modify these values as $\gamma=(0.95\mbox{-}1.05)$, $\gamma=(1.0\mbox{-}1.20)$, and $\gamma=(1.30\mbox{-}1.40)$, respectively. 

On the other hand, the results of PBH-NS events indicate that despite the mismatch of the outcome of the present analysis with GW data, the effect of ellipsoidal-collapse dark matter halo models in the amplification of the merger rate of such binaries is significant. It is also obvious that the merger rate of PBH-NS binaries in both dark matter halo models increases monotonical with the power-law index reaches a maximum at $\gamma=1.7$, and decreases for higher values of $\gamma$.
\section{Conclusions} \label{sec:iv}
In this work, we have calculated the merger rate of compact binaries in dark-matter spikes, which are expected to be structured around SMBHs at the center of galactic halos. For this purpose, we have initially described theoretical models, which suit dark matter spikes. We have also discussed crucial quantities for dark-matter spikes, such as the density profile, concentration parameter, and $M_{\rm SMBH}\mbox{-}\sigma$ relation, which can specify the distribution of dark matter particles in the region of spikes. On the other hand, the strong correlation between the growth of central SMBHs and halo parameters suggests that another quantity called the mass function of SMBHs also plays a prominent role in the present analysis. However, the insufficiency of our knowledge of the exact distribution of dark matter particles in the central regions of galactic halos and the abundance of SMBHs in the Universe may lead to uncertainties in the results. Hence, to manage this uncertainty, we consider three empirical SMBH mass functions to compare their results.

In the following, relying on the PBH scenario and with the assumption that PBHs are capable of contributing to the structure of dark matter, we have discussed the encountering conditions of compact objects such as PBHs and NSs in dark-matter spikes. Consequently, we have calculated the merger rate of compact binaries in a single dark-matter spike for spherical- and ellipsoidal-collapse dark matter halo models. Our results confirm that the merger rate of PBH-PBH binaries within each spike for ellipsoidal-collapse dark matter halo models is slightly higher than that derived from spherical-collapse dark matter halo models. It is concluded that the merger rate of PBH-PBH binaries changes directly with the value of the power-law index $\gamma$. The results obtained from the analysis of PBH-NS events show that the maximum value of the merger rate in each spike takes place around the central SMBH with a mass $M_{\rm SMBH}=10^{7}\, M_{\odot}$. Moreover, it turned out that the merger rate of PBH-NS binaries per spike has a maximum value at $\gamma=1.7$. Also, the results indicate that the merger rate of PBH-NS binaries per spike for ellipsoidal-collapse dark matter halo models is much higher than that derived from spherical-collapse dark matter halo models.

As the main measurable factor in GW detectors, we have calculated the total merger rate of compact binaries in dark-matter spikes around central SMBHs with masses of $M_{\rm SMBH}=(10^{6}\mbox{-}10^{9})M_{\odot}$. As mentioned earlier, to account for possible uncertainties in our analysis, we have used three different mass functions for calculating the total merger rate of compact binaries. Our findings indicate that the merger rate of PBH-PBH binaries in spherical- and ellipsoidal-collapse dark matter halo models has the maximum value at $M_{\rm SMBH}=10^{6}M_{\odot}$ and decreases monotonically with increasing the mass of SMBHs. Also, the results exhibit that the overall merger rate of PBH-PBH binaries for ellipsoidal-collapse dark matter halo models is slightly higher than that obtained from spherical-collapse dark matter halo models, in such a way that in the best case, e.g. at about $\gamma=0.05$, the amplification of the overall merger rate is about $62\%$. This means that in the limit of $r_{\rm sp}\rightarrow 4 r_{\rm s}$, the effect of ellipsoidal-collapse dark matter halo models on the enhancement of the merger rate of PBH-PBH binaries increases. Moreover, among the considered mass functions, the highest, middle, and lowest values of the merger rate of PBH-PBH binaries have been extracted from \cite{2004MNRAS.354.1020S}, \cite{2009MNRAS.400.1451V}, and \cite{2007MNRAS.379..841B} mass functions, respectively. However, the results of the merger rate of PBH-NS binaries were obtained slightly different. Interestingly, the inverse proportionality of the overall merger rate PBH-NS binaries with the mass of SMBHs is not monotonic and has a plateau at $M_{\rm SMBH}=10^{7}M_{\odot}$ in almost all considered models. This may be due to a non-standard abundance of NSs in dark-matter spikes clustered around the central SMBH with of mass $M_{\rm SMBH}=10^{7}M_{\odot}$, which should be validated with informative observational data.

Finally, we have calculated the merger rate of compact binaries according to the power-law index $\gamma$ for ellipsoidal-collapse dark matter halo models and compared them with the corresponding results of spherical-collapse dark matter halo models. To compare our findings with the experimental data of GWs, we have also included the LIGO-Virgo sensitivity band for the merger rate of BH-BH binaries. This task was not possible to do for PBH-NS events, as our results were far from those estimated via the LIGO-Virgo detectors. Our results show that the inclusion of ellipsoidal-collapse dark matter halo models in the calculations of the merger rate of PBH-PBH binaries can reduce the range of the power-law index obtained from spherical-collapse dark matter halo models by $0.1$. This result comes from the fact that the dark-matter spikes in ellipsoidal-collapse halo models are denser (and naturally smaller in radius) than those in spherical-collapse halo models. Additionally, the results show that, unlike PBH-PBH events, the effect of ellipsoidal-collapse dark matter halo models in the amplification of the merger rate of PBH-NS binaries is significant. A maximum at $\gamma=1.7$ is also confirmed for such events.

Although the current analysis gives interesting and reasonable results, the conditions considered in it are not necessarily satisfied and the problem always includes uncertainties. For example, the inaccuracy of the $M_{\rm SMBH}\mbox{-}\sigma$ relation, uncertainties arising from the SMBH mass functions, the incomplete knowledge of the exact distribution of dark matter in the spike regions and the possible disruption of the formed binaries via the surrounding compact objects before their mergers may affect our final results. Hopefully, one can achieve a more detailed insight into the merger event of compact binaries in the vicinity of SMBHs by improving relevant instruments and increasing our understanding of these issues.

\bibliography{sample631}{}
\bibliographystyle{aasjournal}

\end{document}